\documentclass[a4paper, journal,twoside,onecolumn,draftcls]{IEEEtran} %
\usepackage{styles/Notations}

\usepackage[utf8]{inputenc}

\usepackage{calc}

\usepackage{amsmath}				
\usepackage{amssymb}				
\usepackage{amsfonts}			
\usepackage{mathrsfs}			
\usepackage{stmaryrd}			

\usepackage{tikz,pgfplots}
\pgfplotsset{compat=1.13}
\usepackage{pgfplotstable}
\usetikzlibrary{spy,backgrounds}
\usepgfplotslibrary{fillbetween}
\usepackage[ruled,vlined,linesnumbered]{algorithm2e}
\usepackage{algpseudocode}
\usepackage{graphicx}

\usepackage{tabularx}
\usepackage{xcolor}
\usepackage{colortbl}
\usepackage{array}
\usepackage{multirow}

\usepackage{booktabs}

\usepackage{todonotes}

\usepackage{subfigure}

\usepackage{comment}

\usepackage{dsfont} 

\makeatletter
\def\bstctlcite{\@ifnextchar[{\@bstctlcite}{\@bstctlcite[@auxout]}}
\def\@bstctlcite[#1]#2{\@bsphack
  \@for\@citeb:=#2\do{%
    \edef\@citeb{\expandafter\@firstofone\@citeb}%
    \if@filesw\immediate\write\csname #1\endcsname{\string\citation{\@citeb}}\fi}%
  \@esphack}
\makeatother
\newcommand\biblio{%
biblio/strings_all_ref,%
biblio/NewBiblio%
}

\newlength{\mysize}

\renewcommand{\keepcomment}{0}

\usepackage{forloop}
\usepackage{arrayjobx}

\usepackage{tikzscale} 

\usepackage{cite}

\title{Reconstruction of partially sampled multi-band images -- Application to STEM-EELS imaging}
\author{\'Etienne~Monier,~\IEEEmembership{Student Member,~IEEE,} Thomas~Oberlin,~\IEEEmembership{Member,~IEEE,} Nathalie~Brun, \\Marcel~Tenc\'e, Marta~de~Frutos and Nicolas~Dobigeon,~\IEEEmembership{Senior Member,~IEEE,}
\thanks{Part of this work has been funded by CNRS, France, through the Imag'in ARSIS project and the METSA 17A304 program.}
\thanks{E. Monier, Th. Oberlin and N. Dobigeon are with University of Toulouse, IRIT/INP-ENSEEIHT, 31071 Toulouse Cedex 7, France (email: \{etienne.monier, thomas.oberlin, nicolas.dobigeon\}@enseeiht.fr).}
\thanks{N. Brun, M. Tenc\'e and M. de Frutos are with Laboratoire de Physique des Solides, CNRS UMR 8502, Univ. Paris-Sud, Univ. Paris-Saclay, 91405 Orsay Cedex, France (email: \{nathalie.brun, marcel.tence, marta.de-frutos\}@u-psud.fr).}
}

\begin{document}
\bstctlcite{IEEEexample:BSTcontrol}
\maketitle

\begin{abstract}
Electron microscopy has shown to be a very powerful tool to map the chemical nature of samples at various scales down to atomic resolution. However, many samples can not be analyzed with an acceptable signal-to-noise ratio because of the radiation damage induced by the electron beam. This is particularly crucial for electron energy loss spectroscopy (EELS) which acquires spectral-spatial data and requires high beam intensity. Since scanning transmission electron microscopes (STEM) are able to acquire data cubes by scanning the electron probe over the sample and recording a spectrum for each spatial position, it is possible to design the scan pattern and to sample only specific pixels. As a consequence, partial acquisition schemes are now conceivable, provided a reconstruction of the full data cube is conducted as a post-processing step. This paper proposes two reconstruction algorithms for multi-band images acquired by STEM-EELS which exploits the spectral structure and the spatial smoothness of the image. The performance of the proposed schemes is illustrated thanks to experiments conducted on a realistic phantom dataset as well as real EELS spectrum-images.
\end{abstract}


\section{Introduction}
\label{sec:introduction}
\IEEEPARstart{I}{n} an electron microscope, an electron beam is used as the illumination source, which provides a  spatial resolution far superior to that of an optical microscope (nowadays down to $0.1$nm). The present paper considers the scanning transmission electron microscope (STEM), where the beam is focused as a probe which is scanned  over the sample area of interest. STEM is well adapted to the simultaneous acquisition of a variety of signals, both mono and multi-channels, for each probe position. Among the most commonly collected signals are cathodoluminescence, high-angle annular dark field (HAADF) and electron-energy loss spectroscopy (EELS)~\cite{egerton_electron_2011}. 
However, a classical problem encountered is that the  electron beam can induce sample damage, a phenomenon particularly limiting for sensitive materials~\cite{Egerton2004}, e.g., organic materials or molecular structures. Different mechanisms leading to an alteration in the composition and/or structure can be involved depending on the sample composition and the incident electron energy, and the strategy for minimizing damage has to be adapted to each case. Standard acquisition schemes operate sequentially, line-by-line, and thus concentrate electrons in contiguous areas. This scanning mode can accentuate the effects on the data of certain types of damage occurring over an area larger than the beam size: the signal at a given point is affected by the damage originating from the previous acquisitions. To overcome this, common practice is to reduce the electron dose (number of electrons per unit surface) by decreasing the incident beam current or the acquisition time per pixel, which significantly lowers the signal-to-noise ratio (SNR) and the overall image quality. To reduce the cumulative damage on successive pixels, another possibility would be to increase the distance between pixels, but in that case the use of a periodic sampling scheme leads to under-sampling of spatial information. 

Recent works have addressed this issue and proposed random sampling schemes for reducing damage while keeping the best possible SNR and spatial resolution \cite{beche_development_2016}, \cite{stevens_potential_2014}. These methods can also be advantageously used to limit the effect of sample drift by reducing the total acquisition time. Indeed, for a given total electron dose, one can either acquire all pixels at low SNR (small acquisition time per pixel), or \emph{partially} sampling some specific pixels with a higher SNR (i.e., longer acquisition time per pixel). The last method has several advantages as it allows adaptive studies to be envisioned, such as a move detection by considering $J$ successive acquisitions of $100/J$\% of the pixels. Such a random sampling has been implemented on the STEM VG HB 501 microscope in the Laboratoire de Physique des Solides (LPS, Orsay, France). This particular acquisition pattern required the development of scan coils and blanking plates specific controls \cite{RefOrsay2}. Moreover, it also requires computational reconstruction schemes to recover the full data from the partial measurements within a post-processing task.%

Among them, the most common technique which aims at recovering the missing pixels in an image is referred to as inpainting. Numerous techniques have been proposed in the last decade to reconstruct an original image with high accuracy from only partial information. More generally, a wide range of so-called ill-posed inverse problems have been intensively studied through the compressed sensing paradigm. Compressed sensing is a general framework which provides recovery algorithms along with theoretical guarantees for linear under-determined inverse problems, including partial sampling in some basis, or inpainting. The most common key assumptions are to assume that the data is sparse in an appropriate representation space and that the acquisition is random. As a consequence, such algorithms have been implemented in several applied fields such as MRI~\cite{boyer_algorithm_2014}, ultrasonic imaging~\cite{quinsac_bayesian_2011}, astronomy~\cite{bobin_compressed_2008} or tomography in microscopy~\cite{binev_compressed_2012}.
This has been enabled by recent algorithmic advances that efficiently solved high-dimensional optimization problems involving possibly non-smooth or non-convex penalties. Without satisfying the compressed sensing assumptions that are often too restrictive, solving generic but high dimensional image processing problems such as completion~\cite{bertalmio_image_2000}, super-resolution~\cite{zhao_fast_2016} or fusion~\cite{wei_hyperspectral_2015} has been shown to be computationally realizable. 

In particular, the problem of imaging sensitive materials have been recently addressed thanks to partial acquisition schemes coupled with reconstruction methods. These approaches were  based on dictionary learning, sparsity-based or low-rank regularization and applied in transmission electron microscopy (TEM) video~\cite{stevens_applying_2015}, electron tomography~\cite{saghi_reduced-dose_2015} and nuclear magnetic resonance (NMR)~\cite{qu_accelerated_2015}. 
Standard inpainting methods for 2D images are now well-documented and can be applied to two-dimensional HAADF images, for instance by minimizing the Sobolev norm or the total variation or the image (see for instance~\cite{chan_total_2006, chambolle_algorithm_2004}). 
Alternative regularizations include dictionary learning, proposed to solve the sensitive-material problem in HAADF acquisition~\cite{mucke-herzberg_practical_2016,stevens_potential_2014,muecke-herzberg_compressive_2016,european_microscopy_society_compressed_2016}, leading to satisfactory results. 
However, in the case of EELS multi-band images (hereafter referred to as \emph{spectrum-images}), the spatial-spectral structure of the data needs to be taken into account within the reconstruction process. Interestingly, EELS fast acquisition and reconstruction~\cite{stevens_compressive_2016} has raised very few interest compared to its hyperspectral counterpart for Earth observation and remote sensing~\cite{zhang_hyperspectral_2014,chayes_pre-processing_2017,golbabaee_hyperspectral_2012,chen_inpainting_2012}.

This paper considers the reconstruction of EELS spectrum-images from partial acquisition. The proposed reconstruction follows the same lines as the above-referenced techniques, by casting the reconstruction into an inverse problem framework with appropriate regularizations. Compared with the previous works proposed in EELS microscopy, the proposed approach introduces a spatial regularization, chosen to promote smoothly varying images, \emph{and} a spectral regularization to better take into account the nature of EELS data. More precisely, two variants of such a regularization are considered, leading to two different problem formulations. Section~\ref{Section1:InversePb} briefly presents the microscopy material and data. The direct problem, its inverse counterpart and the proposed regularizations are also introduced. The two proposed approaches are then stated in Section~\ref{Sec:ProposedMethods}, while Section~\ref{Section:Implementation} is devoted to their algorithmic implementations. Both approaches are compared and evaluated through numerical simulations conducted on simulated data with controlled ground-truth in Section~\ref{SectionSimulation}, while Section~\ref{SectionRealData} reports experiments on a real dataset acquired by the STEM VG HB 501 microscope operated by LPS. Section~\ref{SectionConclusion} finally concludes this work.

\section{Problem formulation}\label{Section1:InversePb}

\subsection{STEM with variable spatial sampling}

Since this paper is specifically focused on STEM-EELS microscopy, the following paragraph briefly describes this imaging modality (see~\cite{egerton_electron_2011} for a more detailed description). For that purpose, an overall scheme is depicted in Figure~\ref{Fig1:STEMmicro}.
The microscope is based on an electron beam which is emitted by an electron gun. This beam is focused on a sample local zone thanks to magnetic lenses. Then, for each probe position, two signals are recorded in parallel, namely,
\begin{itemize}
\item HAADF 2D imaging: at each sample pixel a real value is measured, depending on the amount of transmitted electrons deviated from the optical axis at a relatively high angle;
\item EELS spectrum-imaging: at each sample pixel an energy-loss spectrum is acquired, produced by inelastically scattered electrons.
\end{itemize}  

\begin{figure}[ht]
\centering
\resizebox{1.1\myfigurewidth}{!}{
\newcommand{\Colors}{{%
"FF0000",
"FF7F00",
"FFFF00",
"00FF00",
"0000FF",
"4B0082",
"8B00FF",
}}

\begin{tikzpicture}[]

\usetikzlibrary{automata,positioning}

\newcounter{index}
\setcounter{index}{0}

\usetikzlibrary{decorations.markings}
\usetikzlibrary{intersections}
\usetikzlibrary{calc}

\tikzset{arrow data/.style 2 args={%
      decoration={%
         markings,
         mark=at position #1 with {\arrow{#2}}},
         postaction={decorate}
      }%
}

\tikzset{
    partial ellipse/.style args={#1:#2:#3}{
        insert path={+ (#1:#3) arc (#1:#2:#3)}
    }
}

\fill [fill = orange!40](0,1) -- (-1.8,4) -- (1.8,4) -- cycle;
\draw [draw = orange!80, thick, arrow data={0.5}{>} ](0,1) -- (-1.8,4);
\draw [draw = orange!80, thick, arrow data={0.5}{>} ](0,1) -- (1.8,4);

\fill [fill = orange!40](-1.8,4) -- (1.8,4) -- (-1.8,8) -- (1.8,8) -- cycle;
\draw [draw = orange!80, thick, arrow data={0.5}{>}](-1.8,4) -- (0,6);
\draw [draw = orange!80, thick, arrow data={0.5}{>}](0,6) -- (1.8,8);
\draw [draw = orange!80, thick, arrow data={0.5}{>}](1.8,4) -- (0,6);
\draw [draw = orange!80, thick, arrow data={0.5}{>}](0,6) -- (-1.8,8);

\fill [fill=orange!40] (-1.8,8) --(1.8,8) --(0,12) --cycle;
\draw [draw = orange!80, thick, arrow data={0.3}{>}] (-1.8,8)--(0,12);
\draw [draw = orange!80, thick, arrow data={0.3}{>}] (1.8,8)--(0,12);


\coordinate (A) at (0,12.25);
\fill [fill = yellow!40!orange, opacity = 0.4] (A) -- (-1.4,14) -- (-0.6,14) -- cycle;
\fill [fill = yellow!40!orange, opacity = 0.4] (A) -- (1.4,14) -- (0.6,14) -- cycle;
\draw [draw = yellow!40!orange, thick, arrow data={0.6}{>}] (A) -- (-1.4,14);
\draw [draw = yellow!40!orange, thick, arrow data={0.7}{>}] (A) -- (-0.6,14);
\draw [draw = yellow!40!orange, thick, arrow data={0.6}{>}] (A) -- (1.4,14);
\draw [draw = yellow!40!orange, thick, arrow data={0.7}{>}] (A) -- (0.6,14);
\filldraw [xshift = -1cm] (-0.4,14) rectangle (0.4,14.1);\filldraw [xshift = 1cm] (-0.4,14) rectangle (0.4,14.1); 

\fill [fill = orange!40] (A) -- (0.9428,15) -- (-0.9428,15) -- cycle;
\draw [draw = orange!80, thick, arrow data={0.8}{>}] (A) --(0.9428,15);
\draw [draw = orange!80, thick, arrow data={0.8}{>}] (A) --(-0.9428,15);

\draw [draw = orange!80, very thick, arrow data={0.8}{>}] (A) --( 0,16);

\filldraw [fill = black!20] (-0.25,0) -- (0.25,0) -- (0,1) -- cycle;							
\filldraw [fill = gray] (0,4) ellipse (2 and 0.25); 												
\filldraw [fill = gray, yshift = 4cm] (0,4) ellipse (2 and 0.25); 						
\filldraw [fill = black!30] (-1.8,12) rectangle (1.8,12.25);														

\filldraw [xshift = -0.55cm] (-0.5,15) rectangle (0.5,15.05);\filldraw [xshift = 0.55cm] (-0.5,15) rectangle (0.5,15.05); 

\foreach \x in {-0.4,-0.2857,...,0.4}{
		\pgfmathsetmacro{\thecurrentcolor}{\Colors[\value{index}]}
        \definecolor{currentcolor}{HTML}{\thecurrentcolor}
		\fill [shading = axis, left color=white, right color=currentcolor,shading angle=135] (\x,16) -- ($(\x ,16) + (0.1143,0)$) to [in = 180, out = 90]($(1.4,17) + (0,0.4-\x) - (0,0.1143)$) -- ($(1.4,17) + (0,0.4-\x) $) to [in = 90, out = 180] (\x,16);
		
		\fill [fill = currentcolor, opacity = 0.4] ($(1.4,17) + (0,0.4-\x) - (0,0.1143)$) -- ($(1.4,17) + (0,0.4-\x) - (0,0.1143) + (2,0)$)  --  ($(1.4,17) + (0,0.4-\x) + (2,0) $) -- ($(1.4,17) + (0,0.4-\x) $) -- cycle;
		\stepcounter{index}
}
\draw [thick] (-0.4,16) -- (0.4,16) to [in = 180, out = 90](1.4,17)-- (1.4,17.8) to [in = 90, out = 180] (-0.4,16);

\filldraw (3.4,17) rectangle (4,17.8);

\draw (-4,0) node[anchor = south] {\Large Gun};
\draw (-4,3.8) node[anchor = south] {\Large Cond. Lens};
\draw (-4,7.8) node[anchor = south] {\Large Obj Lens};
\draw (-4,12) node[anchor = south] {\Large Sample};
\draw (-3,14) node[anchor = south] {\Large HAADF detector};
\draw (-2.5,16.5) node[anchor = south] {\Large EELS spectrometer};
\draw (3.8,15.5) node[anchor = south, text width=2.5cm,align=center] {\Large Scintillator and camera};

\end{tikzpicture}
}
\caption{STEM principle. The electron gun emits an electron beam that is focused on the sample. This interaction is measured by different detectors.}
\label{Fig1:STEMmicro}
\end{figure}

\begin{figure}[ht]
\centering
\resizebox{\columnwidth}{!}{
\input{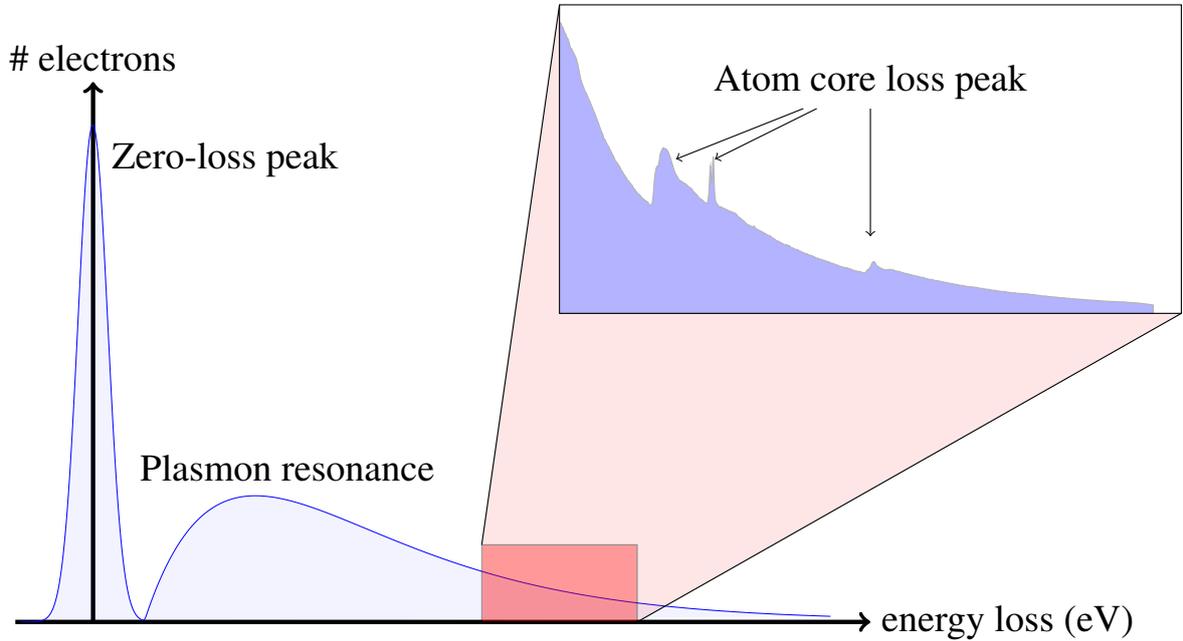}
}
\caption{A typical EELS spectrum.}
\label{Fig2:EELS_curves}
\end{figure}

The present paper focuses on EELS data. A typical EELS spectrum is shown in Figure~\ref{Fig2:EELS_curves}. This spectrum is mainly composed of i) a zero-loss peak corresponding to the electrons that did not lose any detectable energy by interacting with the sample and ii) a plasmon peak corresponding to collective electron excitations. At higher energy loss, the decreasing slope reveals edges giving information about the nature of sample atoms through inner shell excitation. Such a spectrum is acquired for each spatial position of the imaged scene. Thus, multivariate data analysis techniques can provide a wide variety of spatial maps corresponding to chemical and physical properties of the sample.

\subsection{Direct and inverse problems}
\label{subsec:direct_inverse_problems}
Let \Y\ denote the $\Nb\times \Np$ matrix that would correspond to the full EELS data composed of $\Np$ pixels in  \Nb\ channels.
As explained previously, acquiring the whole data cube is not always possible because of possible damaging effects the electron dose could have on a sensitive sample. To avoid this, the scene of interest can be sampled at only some given spatial locations. Note that this spatial subsampling scheme is not accompanied by any spectral subsampling, since, at each spatial position, the EELS spectrometer separates simultaneously all electron loss energies, leading to the acquisition of the whole EELS spectrum. To summarize, the full spectra are acquired at \Ns\ among \Np\ available spatial positions, resulting in the spatial subsampling ratio $\r = \Ns/\Np$. The corresponding index set of the \Ns\ acquired pixels  and the matrix of the measurements are denoted \I\ and $\Y_{\I}$, respectively, where $\Y_{\I}$ is the $\Nb \times \Ns$ matrix gathering the $\Ns$ columns of $\Y$ indexed by $\I$. 

%

The reconstruction problem consists in recovering a full (and possibly denoised) $\Nb\times \Np$ spectrum-image ${\tX}$ from the acquired data $\Y_{\I}$. This reconstruction task is an ill-posed problem and can be addressed by minimizing the criterion
\begin{equation}
\label{eq:min_problem}
\hat{\tX} \in \arg\min_{\tX} \mathcal{L}(\tX,\Y_{\I}) + \lambda \phi(\tX) + \mu \psi(\tX)
\end{equation} 
where $\mathcal{L}(\cdot,\Y_{\I})$ is the data fidelity term and $\phi(\cdot)$ and $\psi(\cdot)$ are spatial and spectral regularizations with corresponding hyperparameters $\lambda$ and $\mu$, respectively.

\subsection{A priori information and regularization}
\label{subsec:regularizations}
As explained previously, recovering the full spectrum-image $\tX$ from the measurements $\Y_{\I}$ is an ill-posed problem, thus requiring appropriate regularizations. This work exploits two different kinds of intrinsic information shared by EELS data, namely spatial smoothness and spectral low-rankness, discussed in what follows.

\subsubsection{Spatial regularizations}

Classical spatial regularizations used in image restoration usually rely on the image gradient $\tX \mathbf{D}$, where $\mathbf{D}$ is the spatial discrete gradient operator applied in each channel independently. This gradient can be minimized with respect to (w.r.t.) its $\ell_2$-norm to promote smoothly varying image or w.r.t. its $\ell_1$-norm when considering total variation (TV) to preserve piecewise-constant content. When dealing with multi-band images, spatial regularizations can be extended to promote specific behavior across channels, as the vector TV~\cite{VectorTV} or the more recent collaborative TV~\cite{Adesso_Condat_2017}. In the applicative context considered in this work, the EELS images are expected to be spatially smooth since the target spatial resolution is relatively low compared to the atomic resolution. As a consequence, the energy of the spatial gradient, which enforces spatial smoothness in each band, will be considered.

\subsubsection{Spectral regularizations}

Multiband images encountered in numerous imaging modalities are known to be highly spectrally correlated and, in most cases, obey a low-rank property. A physically-motivated instance of this property arises when analyzing multi-band images under the unmixing paradigm, which assumes that each measured pixel spectrum can be approximated by a mixture of elementary spectra~\cite{dobigeon_linear_2016,dobigeon_spectral_2012,bioucas-dias_hyperspectral_2012-1}. However, promoting the low-rank structure of the spectrum-image $\tX$ would need to minimize the rank of $\tX$, which is a NP-hard problem. One alternative consists in penalizing its nuclear norm $\left\|\tX\right\|_*$, defined as the $\ell_1$-norm of its singular values. This popular convex relaxation leads to a tractable convex problem~\cite{recht_guaranteed_2010}.

This work will also consider an alternative spectral regularization based on a subspace constrained formulation. Similarly to the strategy already followed in \cite{Wei_IEEE_JSTSP_2015} and \cite{Wei_IEEE_Trans_IP_2015} for multi-band image fusion, the main idea is to estimate beforehand the linear subspace where the pixels of the spectrum-image live, and to reconstruct the whole image in this subspace. More precisely, the subspace of interest is first estimated by conducting a principal component analysis (PCA) of the observed measurement. Then, the minimization problem \eqref{eq:min_problem} is reformulated for the projection of the spectrum-image onto the first principal components. The variances of the principal components are additionally exploited to define an appropriate weighted spectral regularization.

\section{Proposed methods}\label{Sec:ProposedMethods}

The spatial and spectral regularizations described previously lead to two different variational formulations for the EELS image reconstruction problem, which are detailed in the present section.

\subsection{The \SNNFull\ approach}

Given the forward model discussed in Section \ref{subsec:direct_inverse_problems} and the expected spatial and spectral characteristics of the reconstructed EELS spectrum-image detailed in Section \ref{subsec:regularizations}, the first reconstruction method consists in solving the following optimization problem
\begin{equation}\begin{aligned}
\hat{\tX} = \operatornamewithlimits{argmin}_{\tX\in\mathbb{R}^{\Nb\times\Np}}\frac{1}{2}\left\|\Y_\I-\tX_\I\right\|_\mathrm{F}^2 
&+ \frac{\lSNN}{2}||\tX \mathbf{D}||_\mathrm{F}^2\\
&+ \mSNN ||\tX||_*.
\label{eq:SNN}
\end{aligned}\end{equation}
This optimization problem, referred to as \SNNFull\ (\SNN), relies on two hyperparameters $\lSNN$ and $\mSNN$ which adjust the weights of the spatial and spectral regularizations, respectively. The choice of these hyperparameters is discussed in Appendix \ref{app:adjusting_SNN_parameters}.


\subsection{The \SSSFull\ approach}\label{SubSec:3S}
To promote the low-rank property of the reconstructed EELS spectrum-image, the \SNN\ approach introduced in the previous paragraph relies on a soft penalization induced by its nuclear norm. Conversely, the \SSSFull\ (\SSS) approach described in what follows imposes this property through a hard constraint. More precisely, the image to be recovered is assumed to write $\tX=\H\S$ where $\H$ is a $\Nb \times \Nb$ orthonormal matrix defining the data principal component basis and $\S= \left[\mathbf{s}_1,\ldots,\mathbf{s}_{\Np}\right]$ is a $\Nb \times \Np$ matrix which gathers the representation coefficients of the spectrum-pixels in this basis. In this work, the basis $\H$ is supposed to be estimated beforehand by conducting a principal component analysis (PCA) of the observed pixels $\Y_{\I}$. Note that, as discussed in Section \ref{sec:introduction}, a partial spatial sampling of the scene results in a higher SNR than the one obtained with a conventional sampling. Thus, the first principal components of highest energy are expected to span a reliable estimate of the actual signal subspace (assumed to be of dimension $R_{\mathrm{true}}$). 

Given this decomposition, the reconstruction of the spectrum-image $\tX$ can be formulated directly into the principal component basis and boils down to estimating the $\Nb\times\Np$ coefficient matrix $\S$. Thus the quadratic data-fitting term $\left\|\Y_{\I} - \tX_{\I}\right\|_{\mathrm{F}}^2$ already used in the \SNN\ criterion \eqref{eq:SNN} can be replaced by $\left\|\Y_{\I} - \H\S_{\I}\right\|_{\mathrm{F}}^2$ or equivalently, since $\H$ is orthogonal, by $\left\|\H^T\Y_{\I} - \S_{\I}\right\|_{\mathrm{F}}^2$. Similarly, the spatial smoothness promoting term $\left\|\tX\mathbf{D}\right\|^2_{\mathrm{F}}$ in \eqref{eq:SNN} can be rewritten with respect to the representation vectors $\left\|\S\mathbf{D}\right\|^2_{\mathrm{F}}$.

Moreover, when the  eigenvectors $\mathbf{h}_1,\ldots,\mathbf{h}_{\Nb}$ identified by PCA and composing the columns of $\H$ are ordered with respect to eigenvalues sorted in decreasing order, the corresponding representation vectors $\S_{1,:},\ldots,\S_{\Nb,:}$ are expected to be of decreasing energy magnitudes, where $\S_{\indb,:}$ stands for the $\indb$th row of $\S$. In particular, if the pixel spectra lie into a subspace of dimension  ${\rhoM}_{\mathrm{true}}$ with ${\rhoM}_{\mathrm{true}}\leq \Nb$, the squared norm  $\left\|\S_{\indb,:}\right\|_2^2$ of the irrelevant representation vectors is expected to be close to $0$ for $\indb \geq {\rhoM}_{\mathrm{true}}$.  This suggests a weighted penalization of the form $\sum_{\indb=1}^{\Nb} \w_{\indb} \left\|\S_{\indb,:}\right\|_2^2$ with increasing weights $(\w_{\indb})_{\indb=1,\ldots,\Nb}$. The design of the weights is discussed in Appendix~\ref{Appendix:Weights}. 

Finally, the proposed \SSS\ approach consists in solving the following optimization problem
\begin{align}
\hat{\S} &= \underset{\S\in\mathbb{R}^{\Nb\times\Np}}{\arg\min} \frac{1}{2\Nb}\left\|\S \mathbf{D}\right\|_\mathrm{F}^2 + \frac{\lSSS}{2}\sum_{\indb=1}^{\Nb} \w_{\indb} \left\|\S_{\indb,:}\right\|_2^2 \nonumber \\
&\textrm{s.t.}\quad \frac{1}{\Nb}\left\|\H^T\Y_{\I(n)}-\S_{\mathcal{I}(n)}\right\|^2_2\leq\hat{\tsig}^2,\ \forall n\in\llbracket 1, \Ns \rrbracket \label{eq:3S_OP_2}
\end{align}
where $\lSSS$ is a parameter adjusting the relative impact of the spatial and spectral regularizations. In \eqref{eq:3S_OP_2}, the data-fitting term $\mathcal{L}(\cdot,\Y_{\I})$ in \eqref{eq:min_problem} is converted into a constraint, since the squared Euclidean distance between the observations and the solution is expected to be bounded by the noise variance $\sigma^2$. 

In practice, as detailed in Appendices~\ref{Appendix:Weights} and \ref{Appendix:Eigenvalues}, an estimate $\hat{\tsig}^2$ of the noise variance  $\sigma^2$ can be derived from an eigen-analysis of the empirical covariance matrix of the observations. Simultaneously, this analysis shows that the weights $\w_{\indb}$ can be chosen to be infinity for $\indb\geq \rhoM+1$ where $\rhoM$ is an estimate of the actual dimension ${\rhoM}_{\mathrm{true}}$ ($\rhoM\leq\Nb$). This rule systematically implies $\S_{\indb,:}$ to be the null vector for $\indb\geq \rhoM+1$. In other words, the optimization problem \eqref{eq:3S_OP_2} can be equivalently rewritten with respect to a $\rhoM\times\Np$ matrix $\S_{1:\rhoM,:}$ and where all the occurrences of $\Nb$ are replaced by $\rhoM$. The reconstructed image is finally defined as\footnote{To lighten the writing, despite a slight abuse of notations, the underscripts $\cdot_{1:\rhoM,:}$ and $\cdot_{1:\rhoM}$ will be omitted in the sequel of the paper.} $\hat{\tX}=\H_{1:\rhoM}\S_{1:\rhoM,:}$ where $\H_{1:\rhoM}$ stands for the $\Nb\times \rhoM$ matrix composed of the first $\rhoM$ columns of $\H$. The advantages of formulating the reconstruction task as described above are twofold. First, it explicitly imposes a low-rank decomposition of the spectrum-image to be recovered. Second, as discussed in the following implementation section, it reduces the computational cost of the resulting minimization algorithm since $\rhoM$ is expected to be significantly lower than $\Nb$.


\section{Implementation}\label{Section:Implementation}

This section describes the algorithmic implementations derived to solve the optimization problems \eqref{eq:SNN} and \eqref{eq:3S_OP_2}. Both rely on the fast iterative shrinkage thresholding algorithm (FISTA) \cite{beck_fast_2009} briefly recalled in the following paragraph.

\subsection{FISTA general framework}
FISTA solves the generic optimization problem of the form
\begin{equation}
\hat{x} = \arg\min_x f(x) + g(x) \label{eq:GeneralOP}
\end{equation}
where \begin{itemize}
\item $f : \mathbb{R}^p \to \mathbb{R}$ is a convex function, continuously differentiable with $L_{f}$-Lipschitz continuous gradient,
\item $g : \mathbb{R}^p \to \mathbb{R}$ is a convex possibly nonsmooth function.
\end{itemize}
For any $L> L_{f}$, the FISTA algorithm given in Algo.~\ref{Algo1:FISTA} converges toward a solution of~\eqref{eq:GeneralOP}. The specific instances of FISTA for the problems \eqref{eq:SNN} and \eqref{eq:3S_OP_2} under consideration are presented in the following subsections.

\begin{algorithm}\label{Algo1:FISTA}
\caption{FISTA with constant step size \cite{beck_fast_2009}}
\DontPrintSemicolon
\KwSty{Input :}\ {$L> L_{f}$ an upper bound of $L_{f}$}\;
\KwSty{Initialisation :}\ Set $\mathbf{y}^{(1)} = \mathbf{x}^{(0)} \in \mathbb{R}^p$, $\theta^{(1)}=1$, $i=1$
\While{stopping rule not satisfied}{
$\mathbf{x}^{(i)} = \mathrm{prox}_{g/L}\left( \mathbf{y}^{(i)}-\frac{1}{L}\nabla f(\mathbf{y}^{(i)}) \right)$\label{LigneAlgo1} \label{algostep:gradient}\;
$\theta^{(i+1)} = \frac{1}{2} \left(1+\sqrt{1+4(\theta^{(i)})^2}\right)$\label{LigneAlgo2}\;
$\mathbf{y}^{(i+1)} = \mathbf{x}^{(i)} + \left( \frac{\theta^{(i)}-1}{\theta^{(i+1)}} \right) \left(\mathbf{x}^{(i)}-\mathbf{x}^{(i-1)} \right)$\label{LigneAlgo3}\;
$i \leftarrow i+1$\label{LigneAlgo4}
}
\end{algorithm} 

\subsection{Application to \SNN}
To solve  \eqref{eq:SNN}, the \SNN\ method consists in adopting the  following decomposition
\begin{align}
f(\tX) &= \frac{1}{2}\left\|\Y_\I-\tX_\I\right\|_{\mathrm{F}}^2 + \frac{\lSNN}{2}\left\|\tX \mathbf{D}\right\|_{\mathrm{F}}^2\\
g(\tX) &= \mSNN \left\|\tX\right\|_*.
\end{align}
The corresponding gradient of $f(\cdot)$ required in Step \ref{algostep:gradient} of Algo. \ref{Algo1:FISTA} is given by
\begin{equation}
\nabla f(\tX) = (\tX_\I - \Y_\I) - \lSNN \tX \Delta
\end{equation}
where $\Delta = -\mathbf{D}\mathbf{D}^T$ is the discrete spatial Laplacian operator. 
An upper bound of $L_{f}$ is given by
\begin{align*}
&\left\| \nabla f(\tX_1) - \nabla f(\tX_2)\right\|_{\mathrm{F}}\\
&\qquad = \left\| (\tX_{1\I}-\tX_{2\I})  - \lSNN(\tX_1-\tX_2)\Delta \right\|_{\mathrm{F}}\\
&\qquad \leq (\underbrace{1+\lSNN\left\|\Delta\right\|}_{L})\ \left\|\tX_1-\tX_2\right\|_{\mathrm{F}}
\end{align*}
where $\left\|\Delta\right\|$ stands for the spectral norm of the discrete Laplacian, which is 8 in dimension 2. The Lipschitz constant upper bound  is then $L= 1 +8\lSNN$.

Besides, by denoting $\tX=\mathbf{U}\boldsymbol{\Sigma} \mathbf{V}^T$ the singular value decomposition of \tX, where $\boldsymbol{\Sigma} = \mathrm{diag}(\mu_i)$, the proximal operator associated with $g(\cdot)$ is
\begin{equation}
\mathrm{prox}_{g}(\tX) = {\mathbf{U}} \bar{\boldsymbol{\Sigma}} {\mathbf{V}}^T
\end{equation}
where $\bar{\boldsymbol{\Sigma}} = \mathrm{diag}(\bar{\mu}_i)$ contains the soft-thresholded singular values with threshold \mSNN, i.e.\begin{equation}
    \bar{\mu}_i = \mathrm{sgn}(\mu_i)(\mu_i-\mSNN)\mathds{1}_{\mu_i>\mSNN}(\mu_i)
\end{equation}where $\mathrm{sgn}$ is the sign function and where $\mathds{1}$ is the indicator function.

\subsection{Application to \SSS}
When tackling the problem \eqref{eq:3S_OP_2}, the proposed \SSS\ algorithm relies on the following decomposition
\begin{align}
f(\S) &= \frac{1}{2\rhoM}\left\|\S \mathbf{D}\right\|_{\mathrm{F}}^2 + \frac{\lSSS}{2}\sum_{\indb=1}^\rhoM \w_{\indb} \left\|\S_{\indb,:}\right\|_2^2\\
g(\S) &= \sum_{n=1}^{\Ns} \iota_{\mathcal{B}(\H^T\Y_{\I(n)},\sqrt{\rhoM}\hat{\tsig})} (\S_{\I(n)}) \label{eq:3S_g}
\end{align}
where \begin{equation}
\iota_\mathcal{A}(x) = \left\{\begin{array}{lr}
0, & \text{if } x\in\mathcal{A}\\
+\infty, & \text{if } x\notin\mathcal{A}
\end{array}\right.
\end{equation}
is the indicator function related to set $\mathcal{A}$ and  $\mathcal{B}(x_0,r)$ is the closed $\ell_2$-ball of center $x_0$ and radius $r$. The gradient of $f(\cdot)$ is 
\begin{equation}
\nabla f(\S) = -\frac{1}{\rhoM}\S \Delta + \lSSS \W \S\\
\end{equation}
where $\W=\mathrm{diag}\left\{w_1,\ldots,w_{\rhoM}\right\}$ is the diagonal matrix containing the weights. Similar computations as the ones conducted for \SNN\ lead to the upper bound of the Lipschitz constant $L = 8+\lSSS\max_{\indb}\left\{\w_{\indb}\right\}$. Moreover, as shown by \eqref{eq:3S_g}, the function $g(\cdot)$ is separable with respect to the pixel indexes $n\in \llbracket 1,\Ns\rrbracket$. Hence, the proximal operator associated with $g(\cdot)$ consists in projecting $\S_{\I(n)}$ on $\mathcal{B}(\H^T\Y_{\I(n)},\sqrt{\rhoM}\hat{\tsig})$ for all $n\in \llbracket 1,\Ns\rrbracket$.

\subsection{Algorithm complexity}

This paragraph discusses the computational complexity of the \SNN\ and \SSS\ algorithms. When analyzing the generic algorithmic scheme of FISTA in Algo.~\ref{Algo1:FISTA}, the complexity of the two steps in lines~\ref{LigneAlgo1} to~\ref{LigneAlgo4} is needed for both approaches.

First, lines~\ref{LigneAlgo2} and~\ref{LigneAlgo4} are clearly of asymptotic order $\mathcal{O}(1)$. Then, concerning line~\ref{LigneAlgo3}, which consists only in matrix addition, the complexity is of $\mathcal{O}(\Nb\Np)$ and $\mathcal{O}(\rhoM\Np)$ for  \SNN\ and \SSS, respectively. Line~\ref{LigneAlgo1} consists in a gradient descent step, followed by a proximity mapping. The detail is shown in Table~\ref{Tablex:ComplexitySummary}. This study shows that \SNN\ is computationally heavier than \SSS\, mainly because \SNN\ requires an SVD at each iteration, while \SSS\ operates on a matrix of lower dimension since $\rhoM \leq \Nb$.

\begin{table}[h!]
\centering
\caption{Computational complexity of \S2N\ and \SSS.}
\bgroup
\def\arraystretch{1.5}
\begin{tabular}{|b{3cm}|c|c|}
\hline
\rowcolor{black!15}
 &\SNN&\SSS\\
\hline
\hline
Line~\ref{LigneAlgo1} (gradient descent)		&$\mathcal{O}(\Nb\Np)$      &$\mathcal{O}(\rhoM \Np)$\\
\rowcolor{black!5}
Line~\ref{LigneAlgo1} (proximal operator)		&$\mathcal{O}(\Nb\Np^2)$    &$\mathcal{O}(\rhoM \Ns)$\\
Line~\ref{LigneAlgo3}							&$\mathcal{O}(\Nb\Np)$      &$\mathcal{O}(\rhoM\Np)$\\
\rowcolor{black!5}
Lines~\ref{LigneAlgo2} and~\ref{LigneAlgo4}		&$\mathcal{O}(1)$           &$\mathcal{O}(1)$\\
\hline
\rowcolor{black!15}TOTAL						&$\mathcal{O}(\Nb\Np^2)$    &$\mathcal{O}(\rhoM \Np)$\\
\hline
\end{tabular}
\egroup
\label{Tablex:ComplexitySummary}
\end{table}

\section{Simulation results}\label{SectionSimulation}

\subsection{Synthetic datasets}
The performances of the proposed methods are assessed thanks to experiments conducted on synthetic spectrum-images. More precisely, the full spectrum-image $\Y \in\mathbb{R}^{\Nb\times \Np}$ is generated according to
\begin{equation}
\Y= \tX + \mathbf{E}
\end{equation}
where $\tX$ is the noise-free spectrum image and $\mathbf{E}$ is a noise matrix. To mimic realistic EELS acquisitions, the noise-free image has been decomposed as $\tX=\M \A$,  following the so-called linear mixing model that can be used to describe the spatial mapping of materials within an observed sample \cite{dobigeon2012spectral}. The $\Nb\times\Nc$ matrix $\M=\left[\mathbf{m}_1,\ldots,\mathbf{m}_{\Nc}\right]$ gathers $\Nc$ spectra associated with distinct materials (referred to as \emph{endmembers}) and $\A = \left[\mathbf{a}_1,\ldots,\mathbf{a}_{\Np}\right]$ is a $\Nc \times \Np$ matrix which stands for the spatial distribution of the materials in the pixels (referred to as \emph{abundances}).\\

\begin{figure}[h!]
\centering
\resizebox{\columnwidth}{!}{
\input{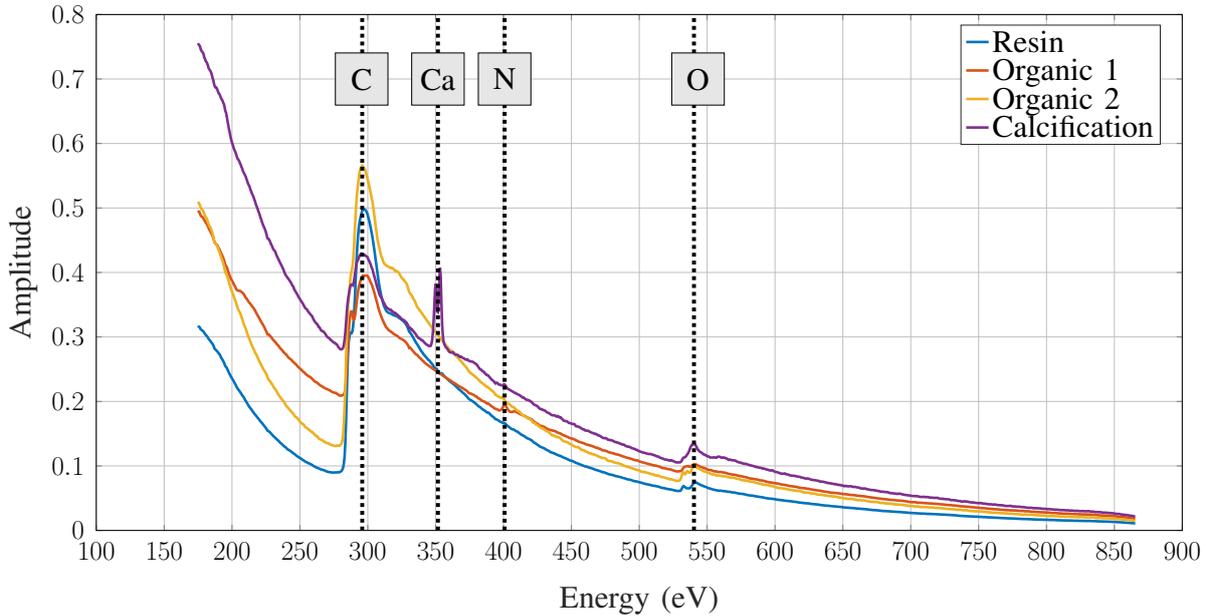}
}
\caption{The $N_c=4$ endmember spectra represented as amplitude vs. energy loss (in eV). The following characteristic thresholds are depicted: carbon (285~eV), calcium (350~eV), nitrogen (400~eV) and oxygen (530~eV).}
\label{Fig6:DataSpectrums}
\end{figure}

\noindent\textbf{Choice of the endmember matrix $\M$:} As it is complicated to simulate whole EELS spectra with different edges and a fine structure for each edge, representative endmember spectra $\mathbf{m}_1,\ldots,\mathbf{m}_{\Nc}$ were directly extracted from a real data set already considered for its medical interest: a section of kidney (biological tissue) embedded in resin and containing calcifications~\cite{RefOrsay36}. For such complex biological samples, the number of endmembers $\Nc$ has to be adjusted on each data set and is typically between $3$ and $5$. Here, the number of extracted endmembers is chosen as Here $\Nc=4$. Endmember extraction is conducted using the vertex component analysis (VCA), a popular algorithm designed for remote sensing hyperspectral images \cite{VCA}, that is now frequently used by the EELS community. The spectra are shown in Figure~\ref{Fig6:DataSpectrums} where four particular \emph{energy thresholds} reveal the presence of chemical elements: carbon ($K$-edge at $285$ eV), calcium ($L_{2,3}$-edge composed of a double peak around $350$ eV), nitrogen ($K$-edge at $400$ eV) and oxygen ($K$-edge at $530$ eV). These components do not correspond to well defined chemicals compounds. Nevertheless, for simplicity, in the following, the endmembers will be related to particular materials and designed as calcification (with Ca $L_{2,3}$-edge), resin, organic 1 (with N-$K$ edge) and organic 2. The number of bands (corresponding to  energy channels of the spectrometer) is $\Nb = 1337$.\\

\begin{figure}[h!]
\centering
\subfigure[Resin]{\includegraphics[width=0.3\columnwidth]{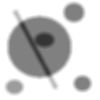} }\hspace{0.1\columnwidth}
\subfigure[Organic 1]{\includegraphics[width=0.3\columnwidth]{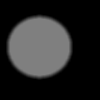} }
\subfigure[Organic 2]{\includegraphics[width=0.3\columnwidth]{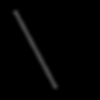} }\hspace{0.1\columnwidth}
\subfigure[Calcification]{\includegraphics[width=0.3\columnwidth]{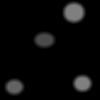} }
\caption{The abundance maps used to generate the synthetic data. A zero-valued (resp. one-valued) abundance coefficient appear in black (resp. white), which corresponds to the absence (resp. presence) of the corresponding endmember.}
\label{Fig5:DataMaps}
\end{figure}

\noindent\textbf{Choice of the abundance matrix $\A$:} The coefficient $a_{kp}$ gives the proportion of the $k$th endmember in the $p$th pixel. To ensure a comprehensive additive description of the spectrum-image in terms of the $\Nc$ materials introduced above, these coefficients are assumed to be nonnegative and subject to sum-to-one constraint for each pixel, i.e.,
\begin{align}
 a_{kp} \geq 0,& \ \  \forall p\in \llbracket 1, \Np \rrbracket , \forall k\in \llbracket 1, \Nc \rrbracket,\\
\sum_{k=1}^{\Nc} a_{kp} = 1,&\ \  \forall p\in \llbracket 1, \Np \rrbracket.
\end{align}
According to these constraints, four abundance maps $\A_{k,:}=\left[a_{k1},\ldots,a_{k\Np}\right]$ ($k\in \llbracket 1, \Nc \rrbracket$) represented in Fig.~\ref{Fig5:DataMaps}  have been designed to define the spatial distribution of the different materials in the sample. In these experiments, the spatial maps are of size  $100\times100$ pixels, which corresponds to $\Np = 10000$.\\

\noindent\textbf{Generation of the noise matrix $\mathbf{E}$:} The components of the noise matrix $\mathbf{E}$ are independently and identically randomly generated according to a centered Gaussian distribution. The noise variance has been adjusted to reach realistic signal-to-noise ratios (SNRs) that will be specified later.

\subsection{Performance w.r.t. the regularization parameters}\label{SubSec:Hyper-MSE}
We first evaluate the impact of the regularization parameters ($\lSNN,\mSNN$) and $\lSSS$ on the quality of the spectrum-image reconstructed by the two proposed algorithms \SSS\ and \SNN\ for a noise level of $\mathrm{SNR}=25$dB and a sampled pixel ratio $r=\Ns/\Np=0.2$. The considered figure-of-merit is the normalized mean square error (NMSE) 
\begin{equation}
\label{eq:NMSE}
\mathrm{NMSE}(\tX,\hat{\tX}) = \frac{||\hat{\tX} - \tX||_{\mathrm{F}}^2}{||\tX||_{\mathrm{F}}^2}
\end{equation}
associated with the reconstructed image $\hat{\tX}$ obtained for various parameter values.

\begin{figure}[h!]
\centering
\resizebox{\columnwidth}{!}{
{\Large
\input{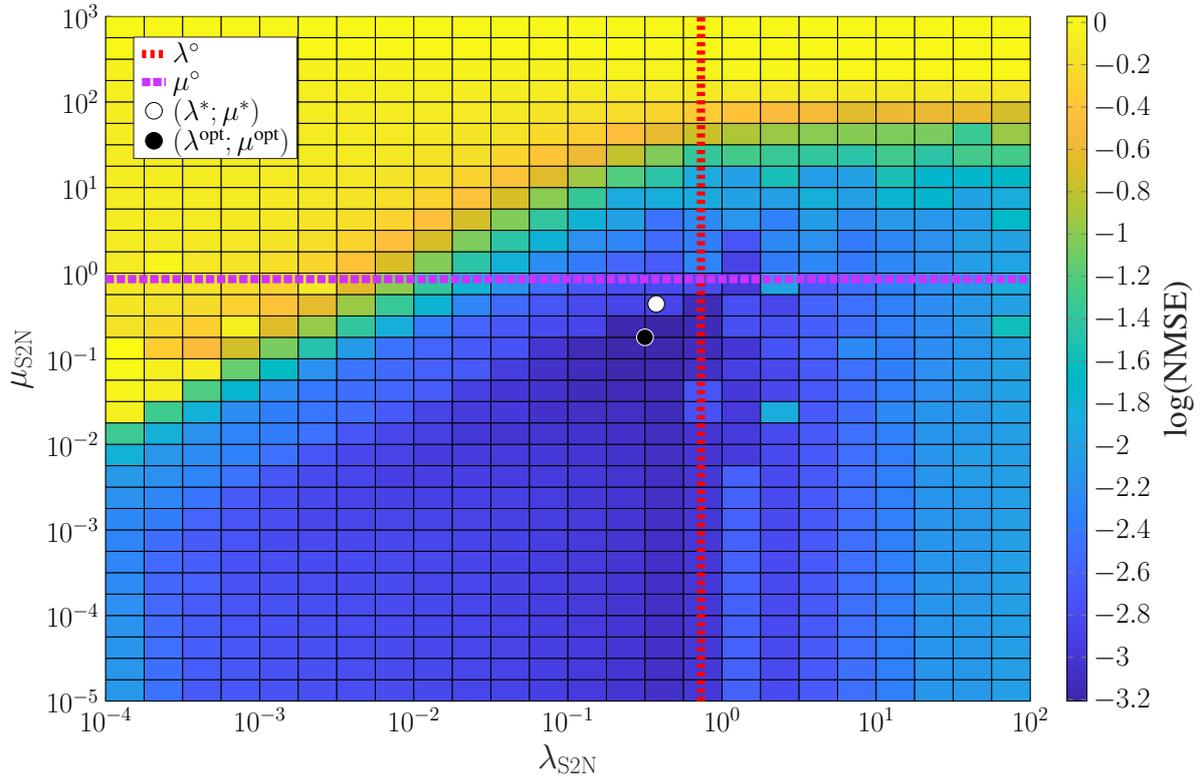}
}
}
\caption{\SNN\: NMSE as a function of $(\lSNN,\mSNN)$. The white marker locates to the parameters $(\lSNN^{*},\mSNN^*)$ tuned following the procedure described in Appendix \ref{app:adjusting_SNN_parameters}. The vertical and horizontal lines correspond to the intermediate values $\lSNN=\lSNN^{\circ}$ and $\mSNN=\mSNN^\circ$ retrieved during this procedure. The black marker shows the optimal parameter set $(\lSNN^{\mathrm{opt}},\mSNN^{\mathrm{opt}})$ leading to the minimal MSE value on the grid.}
\label{Fig8:SNN_MSE_Lambda_Mu}
\end{figure}

Figure~\ref{Fig8:SNN_MSE_Lambda_Mu} depicts the performance results of the \SNN\ algorithm as a function of the regularization parameters $(\lSNN,\mSNN)$ on a given grid. In addition, some specific values are reported in Table~\ref{Table 2 : MSE_Hyper}. In particular, the minimum NMSE obtained on the grid is located with a black dot whereas the parameter values obtained by the method described in Appendix~\ref{SubSec:empirical-search} is given with a white dot. The vertical and horizontal lines corresponds to the intermediate values  $\lSNN^\circ$ and $\mSNN^\circ$ obtained during the procedure by independently adjusting the spatial or spectral regularizations while removing the other. As expected, this figure shows that the optimal NMSE is reached for  non-zero values for both parameters, demonstrating that both spatial and spectral regularizations are needed. Indeed, extreme values of $\lSNN$ give too smooth images or too rough images. Similarly, high values of $\mSNN$ lead to trivial rank-one images whereas a too low value does not spectrally regularize the image. Moreover the intermediate $\lSNN^\circ$ and $\mSNN^\circ$ values recovered by adjusting the regularization separately tend to over-estimate each regularization compared to the optimal ones. This  behavior was expected, and scaling these values as explained in Appendix~\ref{SubSec:empirical-search} leads to a reasonably efficient tuning. Note however that even though those values are close to the optimal ones, the NMSE values reported in Table~\ref{Table 2 : MSE_Hyper} show that the corresponding NMSE is about twice worse: the algorithm seems quite sensitive to the parameters.

\begin{table}[h!]
\centering
\caption{\SNN\ and \SSS\: NMSE for particular values of the regularization parameters.}
\bgroup
\def\arraystretch{1.7}%
\begin{tabular}{|>{\centering\arraybackslash}m{2.5cm}|>{\centering\arraybackslash}m{2.5cm}|>{\centering\arraybackslash}m{1.5cm}|}
\hline
\rowcolor{black!20}
Parameters				& Values	&NMSE\\
\hline
\hline
\multirow{ 2}{*}{(\lSNN,\mSNN)}   & $(\lSNN^{\mathrm{opt}},\mSNN^{\mathrm{opt}})$	&$6.209\ 10^{-4}$\\
 	&$(\lSNN^*, \mSNN^*) $		&$1.521\ 10^{-3}$\\
 \hline
\lSSS		&$1$						&$2.494\ 10^{-4}$\\
\hline
\end{tabular}
\egroup
\label{Table 2 : MSE_Hyper}
\end{table}

Thanks to its constrained formulation, the \SSS\ algorithm requires to adjust only one regularization parameter, namely $\lSSS$, which balances the relative contributions of the  spectral and spatial regularizations. When $\lSSS$ is too small (resp. large), the spatial regularization becomes preponderant (resp. negligible), which leads to an over- (resp. under-) smoothed image. Experiments (not reported in this paper for brevity) have shown that the \SSS\ is not very sensitive to the tuning of the hyperparameter and that choosing $\lSSS=1$ consistently leads to satisfactory results (the corresponding NMSE is reported in Table~\ref{Table 2 : MSE_Hyper}). This value will be subsequently used in the experiments reported in this section.


\subsection{Performance w.r.t. noise level and sampling ratio}
\begin{figure}[h!]
\centering
\resizebox{0.8\columnwidth}{!}{
\begin{tikzpicture}

\pgfplotstableread{
SNR Mean Sup Low
19.0000000000 0.0003098895 0.0003141262 0.0003056528
21.0000000000 0.0002946734 0.0002959200 0.0002934268
23.0000000000 0.0002793770 0.0002805869 0.0002781670
25.0000000000 0.0002669063 0.0002675667 0.0002662458
27.0000000000 0.0002585467 0.0002594032 0.0002576901
29.0000000000 0.0002525841 0.0002534295 0.0002517387
}\pixonetable

\pgfplotstableread{
SNR Mean Sup Low
19.0000000000 0.0001607866 0.0001619001 0.0001596732
21.0000000000 0.0001470987 0.0001479504 0.0001462470
23.0000000000 0.0001356802 0.0001364536 0.0001349068
25.0000000000 0.0001271508 0.0001275902 0.0001267113
27.0000000000 0.0001202579 0.0001205534 0.0001199624
29.0000000000 0.0001158253 0.0001161259 0.0001155246
}\pixtwotable

\pgfplotstableread{
SNR Mean Sup Low
19.0000000000 0.0000964617 0.0000968294 0.0000960941
21.0000000000 0.0000829103 0.0000835565 0.0000822641
23.0000000000 0.0000736752 0.0000741392 0.0000732112
25.0000000000 0.0000670064 0.0000673893 0.0000666235
27.0000000000 0.0000627092 0.0000628590 0.0000625593
29.0000000000 0.0000593730 0.0000594920 0.0000592539
}\pixthreetable

\pgfplotstableread{
SNR Mean Sup Low
19.0000000000 0.0039885167 0.0068900845 0.0010869490
21.0000000000 0.0015366018 0.0017723822 0.0013008213
23.0000000000 0.0013272545 0.0019691254 0.0006853837
25.0000000000 0.0010503906 0.0013089202 0.0007918611
27.0000000000 0.0016200434 0.0016227074 0.0016173795
29.0000000000 0.0013038972 0.0013054698 0.0013023245
}\pixonetableSNN

\pgfplotstableread{
SNR Mean Sup Low
19.0000000000 0.0012224620 0.0013403831 0.0011045410
21.0000000000 0.0011447368 0.0013882078 0.0009012658
23.0000000000 0.0008445761 0.0008946618 0.0007944904
25.0000000000 0.0006802453 0.0007014732 0.0006590175
27.0000000000 0.0011502482 0.0013446166 0.0009558798
29.0000000000 0.0004419364 0.0004520551 0.0004318178
}\pixtwotableSNN

\pgfplotstableread{
SNR Mean Sup Low
19.0000000000 0.0015778387 0.0032382786 0.000326011
21.0000000000 0.0019467793 0.0039271395 0.000335809
23.0000000000 0.0008357649 0.0011683945 0.0005031353
25.0000000000 0.0004969984 0.0005154659 0.0004785309
27.0000000000 0.0007314615 0.0010741257 0.0003887973
29.0000000000 0.0003312282 0.0003414979 0.0003209585
}\pixthreetableSNN

\begin{axis}[
  width=0.9\columnwidth,
  xmin=19,
  xmax=29,
  ymin=0.00006,
  ymax=0.01,
  grid,
  axis x line=left,
  axis y line=left,
  xlabel={SNR (dB)},
  ylabel={NMSE},
  ylabel style={align=center},
  legend columns=2, 
        legend style={
            /tikz/column 2/.style={
                column sep=5pt,
            },
        },
  legend style={legend cell align=left, align=left, draw=white!15!black,font=\scriptsize,at={(0.5,1.05)},anchor=south}, 
  legend entries={SSS - $\pixratio= 0.2$, SNN - $\pixratio= 0.2$, 
								SSS -  $\pixratio= 0.3$, SNN - $\pixratio= 0.3$,
                  				SSS -  $\pixratio= 0.4$, SNN - $\pixratio= 0.4$},
 ymode=log,
]

\tikzstyle{mark1}=[mark=triangle,mark size=4pt]
\tikzstyle{mark2}=[mark=square,mark size=4pt]

\tikzstyle{color1}=[blue!80]
\tikzstyle{color2}=[red!80]
\tikzstyle{color3}=[green!80]

\addlegendimage{mark1,color1}
\addlegendimage{mark2,color1}

\addlegendimage{mark1,color2}
\addlegendimage{mark2,color2}

\addlegendimage{mark1,color3}
\addlegendimage{mark2,color3}

\addplot[color1,mark1] table [x index=0, y index=1]{\pixonetable};
\addplot[name path=A, draw=none] table [x index=0, y index=2]{\pixonetable};
\addplot[name path=B, draw=none] table [x index=0, y index=3]{\pixonetable};
\addplot[color1, opacity=0.3] fill between[of=A and B];

\addplot[color2,mark1] table [x index=0, y index=1]{\pixtwotable};
\addplot[name path=A, draw=none] table [x index=0, y index=2]{\pixtwotable};
\addplot[name path=B, draw=none] table [x index=0, y index=3]{\pixtwotable};
\addplot[color2, opacity=0.3] fill between[of=A and B];

\addplot[color3,mark1] table [x index=0, y index=1]{\pixthreetable};
\addplot[name path=A, draw=none] table [x index=0, y index=2]{\pixthreetable};
\addplot[name path=B, draw=none] table [x index=0, y index=3]{\pixthreetable};
\addplot[color3, opacity=0.3] fill between[of=A and B];

\addplot[color1,mark2] table [x index=0, y index=1]{\pixonetableSNN};
\addplot[name path=A, draw=none] table [x index=0, y index=2]{\pixonetableSNN};
\addplot[name path=B, draw=none] table [x index=0, y index=3]{\pixonetableSNN};
\addplot[color1, opacity=0.3] fill between[of=A and B];

\addplot[color2,mark2] table [x index=0, y index=1]{\pixtwotableSNN};
\addplot[name path=A, draw=none] table [x index=0, y index=2]{\pixtwotableSNN};
\addplot[name path=B, draw=none] table [x index=0, y index=3]{\pixtwotableSNN};
\addplot[color2, opacity=0.3] fill between[of=A and B];

\addplot[color3,mark2] table [x index=0, y index=1]{\pixthreetableSNN};
\addplot[name path=A, draw=none] table [x index=0, y index=2]{\pixthreetableSNN};
\addplot[name path=B, draw=none] table [x index=0, y index=3]{\pixthreetableSNN};
\addplot[color3, opacity=0.3] fill between[of=A and B];

\end{axis}
\end{tikzpicture}
}
\caption{Performance of \SNN\ and \SSS\ in term of NMSE as functions of the pixel ratio $r$ and noise level. Colored filling corresponds to standard deviation interval.}
\label{Fig9:PixSigma}
\end{figure}
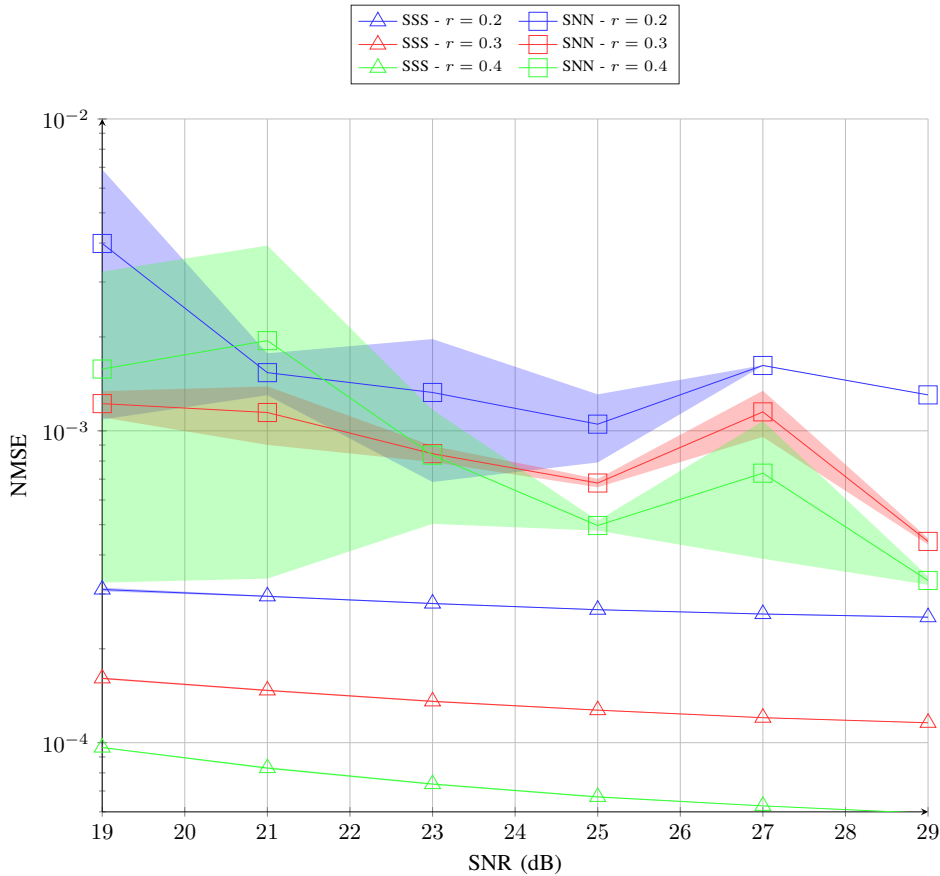

Then, the two algorithms have been  evaluated for various noise levels and pixel ratios $r$. The NMSE have been averaged over $10$ Monte Carlo simulations, for each of which an independent noise matrix $\mathbf{E}$ has been drawn while the sampled pixel mask $\I$ has been kept fixed. The results are shown in Figure~\ref{Fig9:PixSigma}. According to this figure, \SSS\ seems to give good reconstruction results, with smaller NMSE and smaller variances. Conversely, \SNN\ produces higher NMSE (nearly ten times greater than those obtained by \SSS) and higher variances, especially for low SNR values. This difference between both algorithms may have two main explanations. First, as seen previously, the parameters of \SNN\ seem more difficult to be tuned. Second, the nuclear norm used in \SNN\ is known to be a biased proxy for the rank, thus reducing the performance of the reconstruction.

\subsection{Reconstruction vs. denoising w.r.t. an unmixing task}
\label{SubSection3:approachesMetrics}\label{Subsection:Approaches}
Typical acquisition conditions (referred to as protocol $\mathcal{P}_0$) are defined by a sequential sampling ($r=1$) with an acquisition time of $\Delta t=10$ms per pixel. As explained in Section \ref{sec:introduction}, organic samples are easily deteriorated by the electron radiation during the acquisition process. To overcome this issue, the total electron dose should be reduced by adjusting either the ratio $r$ of visited locations or the time spent $\Delta t$ to acquire the spectrum in each spatial location. Two sampling strategies have been envisioned to try to reduce sample damage at a given beam current.

The first protocol, denoted $\mathcal{P}_1$, consists in acquiring the spectra in all the spatial locations ($r=1$), but reducing the acquisition time $\Delta t=2 $ms for each pixel. The resulting decrease of the signal-to-noise ratio (SNR) can be mitigated by a subsequent denoising step. The second acquisition strategy $\mathcal{P}_2$, which motivated the proposed work, consists in acquiring a subset of pixel spectra (i.e., $r \leq 1$) with a same acquisition time $\Delta t =10$ ms (i.e., higher SNR). To compare these two acquisition protocols, two distinct datasets have been generated, corresponding to the same total beam energy $\mathcal{E}$
\begin{itemize}
    \item Protocol $\mathcal{P}_1$: $\Delta t=2$ms, $r = 1$,
    \item Protocol $\mathcal{P}_2$: $\Delta t=10$ms, $r = 0.2$.
\end{itemize}
For each protocol, the noise levels have been adjusted to reach realistic SNR encountered in typical acquisitions for these exposition time $\Delta t$: $\mathrm{SNR}=19$dB and  $\mathrm{SNR}=25$dB for Protocols $\mathcal{P}_1$ and $\mathcal{P}_2$, respectively.

To evaluate the interest of the partial sampling paradigm, the exploitability of the reconstructed images after an acquisition process according to the protocol $\mathcal{P}_2$ is compared with the exploitability of the full image acquired under the experimental protocol $\mathcal{P}_1$ with respect to the ground truth image (referred to as \emph{oracle} in what follows). Moreover, a denoised version of the image acquired according to Protocol $\mathcal{P}_1$ is also considered, where the denoising step consists in combining a thresholded PCA followed by 3D NL-means~\cite{buades2005non}. These $5$ compared images will be referred to as: 
\begin{itemize}
\item Oracle: the noise-free synthetic data,
\item \Fulla: image acquired following $\mathcal{P}_1$,
\item \deFull: denoised version of the image acquired following $\mathcal{P}_1$,
\item \SNN: image reconstructed by \SNN\ after $\mathcal{P}_2$,
\item \SSS: image reconstructed by \SSS\ after $\mathcal{P}_2$.
\end{itemize}

The exploitability of these $5$ images is evaluated with respect to a conventional task frequently conducted when analyzing EELS spectrum-image. Indeed, since experimentalists are rather interested by the composition of the sample, they resort to various \emph{unmixing} techniques elaborated to recover both endmember spectra and abundance maps of interesting components from the spectrum-image \cite{dobigeon2012spectral}. Therefore, on each spectrum-image, the spectra $\mathbf{m}_k$, ${k \in \llbracket 1,\ldots,\Nc \rrbracket}$, have been recovered using the SISAL algorithm~\cite{bioucas2009variable} applied on the acquired data under Protocols $\mathcal{P}_1$ or $\mathcal{P}_2$. The quality of the estimated endmember matrix $\hat{\mathbf{M}}$ is evaluated using the average spectral angle distance (aSAD) defined by \cite{keshava2004distance,sohn2002supervised}  
\begin{equation}
\mathrm{aSAD}(\mathbf{M},\hat{\mathbf{M}}) = \frac{1}{\Nc}\sum_{k=1}^{\Nc}\mathrm{acos}\left(\frac{\langle\mathbf{m}_k, \hat{\mathbf{m}_k}\rangle}{||\mathbf{m}_k||_2\times||\hat{\mathbf{m}_k}||_2}\right).
\end{equation}
The aSAD gives a small value when both actual and estimated spectra are approximately equal or collinear, while different spectra shapes will produce a high aSAD.

Based on the endmember estimates, the abundance maps $\mathbf{A}$ are estimated from the compared images using the SUNSAL algorithm~\cite{bioucas2010alternating}. The relevance of the estimated abundance maps is evaluated by computing the corresponding $\mathrm{NMSE}(\mathbf{A},\hat{\mathbf{A}})$ as defined by \eqref{eq:NMSE}.

In addition to this quantitative assessment in terms of unmixing performance, a qualitative evaluation is conducted by visual inspection of the reconstructed images. Synthetic red-green-blue compositions of the images of interest are generated by selecting 3 specific bands (energy channels) associated with the presence of chemical elements: $b_{\mathrm{red}}=236$ (carbon), $b_{\mathrm{green}}=346$ (calcium) and $b_{\mathrm{blue}}=709$ (oxygen). To ensure fair comparisons between the images, the channels are independently scaled with respect to a dynamic range common for all the images.

The quantitative results are reported in Table~\ref{Table4:Metrics} while the reconstructed images and estimated abundance maps are depicted in Figure~\ref{Fig9:MapsSUNSAL}. Note that the unmixing results obtained from the four acquired, denoised  and reconstructed images are also compared with those that would be obtained directly on the ground truth image $\tX$. These oracle estimates give the most optimistic performance that could be reached when unmixing the denoised or reconstructed images.

\newlength{\mylength}
\setlength{\mylength}{1.8cm}
\begin{table}[ht!]
\centering
\caption{Reconstruction and unmixing performance.}
\bgroup
\def\arraystretch{1.5}%
\begin{tabular}{|>{\centering\arraybackslash}m{1.7cm}|>{\centering\arraybackslash}m{\mylength}|>{\centering\arraybackslash}m{\mylength}|>{\centering\arraybackslash}m{\mylength}|}
\hline
\rowcolor{black!20}
Image	&$\mathrm{NMSE}(\tX,\hat{\tX})$		
&$\mathrm{aSAD}(\mathbf{M},\hat{\mathbf{M}})$	
&$\mathrm{NMSE}(\mathbf{A},\hat{\mathbf{A}})$\\
\hline
\hline
Oracle
&0.00000
&0.08291
&0.32511
\\%
\rowcolor{black!5}\Fulla
&0.01196
&0.12794
&0.52427
\\%
\deFull
&0.00005
&0.12794
&0.51953
\\%
\rowcolor{black!5}\SNN	
&0.00100
&0.11019
&0.48892
\\%
\SSS
&0.00026
&0.11019
&0.40646
\\
\hline
\end{tabular}
\egroup
\label{Table4:Metrics}
\end{table}

\begin{figure*}[hbtp]
\centering
\setlength{\mylength}{2cm}

\bgroup
\def\arraystretch{1.5}%
\begin{tabular}{>{\centering\arraybackslash}m{0.02cm}>{\centering\arraybackslash}m{\mylength}>{\centering\arraybackslash}m{\mylength}>{\centering\arraybackslash}m{\mylength}>{\centering\arraybackslash}m{\mylength}>{\centering\arraybackslash}m{\mylength}>{\centering\arraybackslash}m{\mylength}}
&&Oracle&\Fulla &\deFull&\SNN &\SSS\\
%
& \vfill \vfil True abundance maps
&\includegraphics[width=0.12\textwidth]{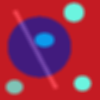}
&\includegraphics[width=0.12\textwidth]{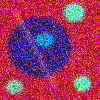}
&\includegraphics[width=0.12\textwidth]{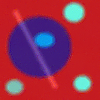}
&\includegraphics[width=0.12\textwidth]{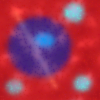}
&\includegraphics[width=0.12\textwidth]{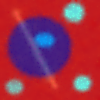}\\
\rotatebox{90}{Resin}
&\includegraphics[width=0.12\textwidth]{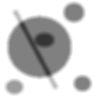}
&\includegraphics[width=0.12\textwidth]{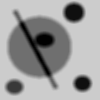}
&\includegraphics[width=0.12\textwidth]{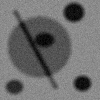}
&\includegraphics[width=0.12\textwidth]{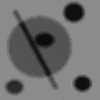}
&\includegraphics[width=0.12\textwidth]{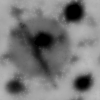}
&\includegraphics[width=0.12\textwidth]{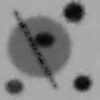}\\
\rotatebox{90}{Organic 1}
&\includegraphics[width=0.12\textwidth]{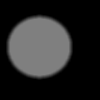}
&\includegraphics[width=0.12\textwidth]{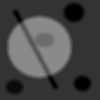}
&\includegraphics[width=0.12\textwidth]{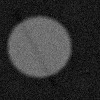}
&\includegraphics[width=0.12\textwidth]{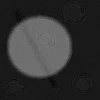}
&\includegraphics[width=0.12\textwidth]{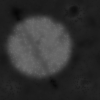}
&\includegraphics[width=0.12\textwidth]{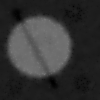}\\
\rotatebox{90}{Organic 2}
&\includegraphics[width=0.12\textwidth]{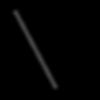}
&\includegraphics[width=0.12\textwidth]{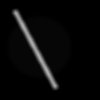}
&\includegraphics[width=0.12\textwidth]{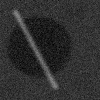}
&\includegraphics[width=0.12\textwidth]{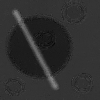}
&\includegraphics[width=0.12\textwidth]{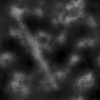}
&\includegraphics[width=0.12\textwidth]{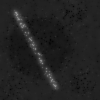}\\
\rotatebox{90}{Calcification}
&\includegraphics[width=0.12\textwidth]{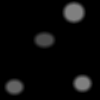}
&\includegraphics[width=0.12\textwidth]{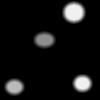}
&\includegraphics[width=0.12\textwidth]{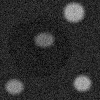}
&\includegraphics[width=0.12\textwidth]{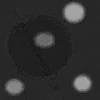}
&\includegraphics[width=0.12\textwidth]{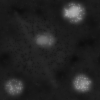}
&\includegraphics[width=0.12\textwidth]{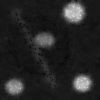}\\
\end{tabular}
\egroup
\caption{Row 1: colored composition of the oracle, acquired, denoised and reconstructed spectrum-images. Rows 2-4: the abundance maps estimated by SUNSAL on the corresponding spectrum-images.}
\label{Fig9:MapsSUNSAL}
\end{figure*}

One can first observe that denoising the image acquired following Protocol $\mathcal{P}_1$ leads to significant lower NMSE for the spectrum-image than those obtained by reconstructing the image acquired following Protocol $\mathcal{P}_2$. The proposed reconstruction algorithms applied after partial sampling seem to be less efficient, despite higher SNR during the acquisition process. However, the performance w.r.t. the unmixing task is in favor of Protocol $\mathcal{P}_2$. Indeed, the endmember spectra are better recovered in the image acquired with longer exposition time than in the image acquired with shorter acquisition time, even after a denoising step. This finding was quite expected for the recovered endmember spectra, since the observed pixels within partial sampling have a higher SNR. Thus the SISAL extraction endmember algorithm can exploit more reliable measured pixel spectra to estimate the material spectral signatures. Moreover, the abundance maps estimated on the image reconstructed by the proposed \SSS\ algorithm achieves a lower NMSE, demonstrating that this reconstructed image can be reliably exploited to spatially map the materials in the spectrum-image.
 
   Qualitatively, visual inspection of the abundance maps depicted in Figure~\ref{Fig9:MapsSUNSAL} shows that the reconstruction techniques produce rougher maps, with some holes corresponding to non-sampled areas. But there is less mixing between the different components, e.g., in particular for Organic 1 and Organic 2 maps recovered on the image reconstructed by \SSS. To summarize, for the same amount of beam energy, partial sampling seems to enable better endmember extraction and spectrum detail recovery than denoising, even though thin spatial  structures (such as Organic 2 in our example) may be spatially mapped with less accuracy.

\section{A real-data example}\label{SectionRealData}

\begin{figure}[h!]
\centering
\subfigure[b][Large view]{\includegraphics[width=0.4\columnwidth]{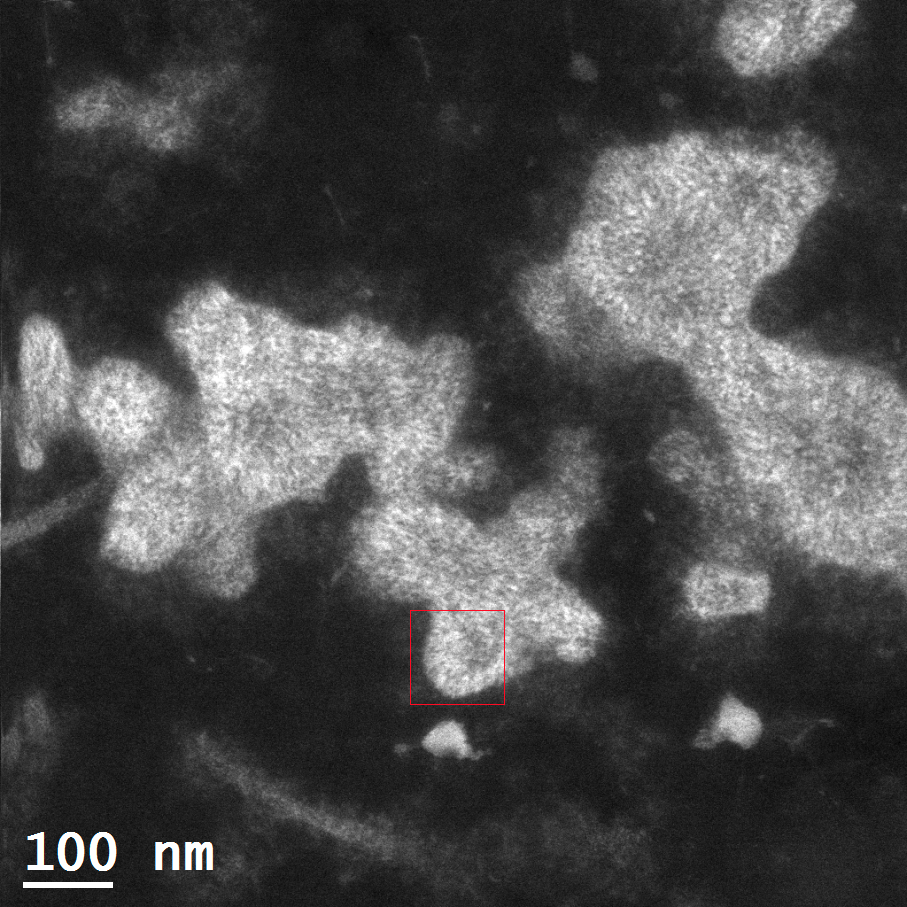} }
\subfigure[b][Area of interest]{\includegraphics[width=0.4\columnwidth]{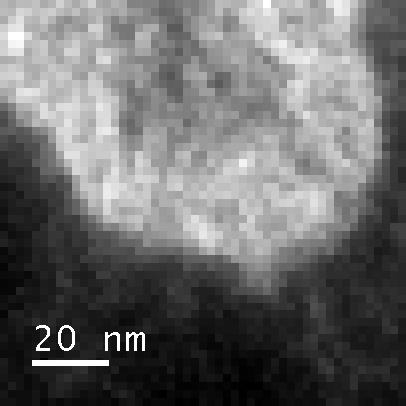} }
\caption{The HAADF images of the sample zone (a) and of the $51 \times 51$ region of interest (b).}
\label{Fig11:HAADFrealdata}
\end{figure}

In this section, the proposed reconstruction methods are applied to a real sample whose gray-scale HAADF image is given in Figure~\ref{Fig11:HAADFrealdata}. This sample has been acquired by the STEM VG HB 501 microscope and equipped with a partial sampling implementation. 

\begin{figure}[h!]
\centering
\subfigure[b][\Fullb]{\includegraphics[width=0.28\columnwidth]{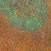} }\hspace{0.01\columnwidth}
\subfigure[b][\Fulla]{\includegraphics[width=0.28\columnwidth]{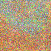} }\hspace{0.01\columnwidth}
\subfigure[b][\Partial]{\includegraphics[width=0.28\columnwidth]{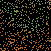} }
\subfigure[b][\deFull]{\includegraphics[width=0.28\columnwidth]{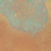} }\hspace{0.01\columnwidth}
\subfigure[b][\SNN]{\includegraphics[width=0.28\columnwidth]{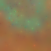} }\hspace{0.01\columnwidth}
\subfigure[b][\SSS]{\includegraphics[width=0.28\columnwidth]{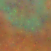} }
\caption{Colored compositions of the real spectrum-images.
}
\label{Fig12:SpimDisplay}
\end{figure}

In the experimental setup, three distinct protocols are considered to image the sample:
\begin{itemize}
    \item Protocol $\mathcal{P}_0$: $\Delta t=10$ms, $r = 1$
    \item Protocol $\mathcal{P}_1$: $\Delta t=2$ms, $r = 1$,
    \item Protocol $\mathcal{P}_2$: $\Delta t=10$ms, $r = 0.2$,
    .    
\end{itemize}
Protocol $\mathcal{P}_0$ corresponds to usual acquisition parameters for this type of samples.
Finally, the images compared in this section are the following
\begin{itemize}
    \item \Fullb: image acquired following $\mathcal{P}_0$,
    \item \Fulla: image acquired following $\mathcal{P}_1$,
    \item \Partial: image acquired following $\mathcal{P}_2$,  
    \item \deFull: denoised version of the image acquired following $\mathcal{P}_1$,
    \item \SNN: image reconstructed by \SNN\ after $\mathcal{P}_2$,
    \item \SSS: image reconstructed by \SSS\ after $\mathcal{P}_2$.
\end{itemize}
The acquired image size is  with $51\times51$ pixels and synthetic colored compositions of these images are represented in Figure~\ref{Fig12:SpimDisplay} where \Fulla\ is used as a reference image to define the color dynamics.

\begin{figure}[hbtp]
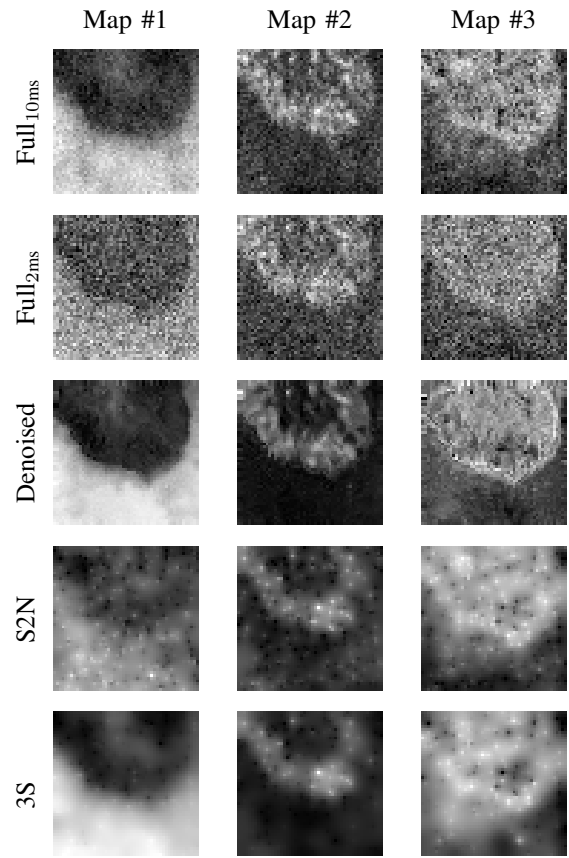

\centering
\setlength{\mylength}{2cm}
\newlength\mylengths
\setlength{\mylengths}{0.12\textwidth}
\newcounter{ct}

\bgroup
\def\arraystretch{1.5}%
\begin{tabular}{>{\centering\arraybackslash}m{0.02cm}>{\centering\arraybackslash}m{\mylength}>{\centering\arraybackslash}m{\mylength}>{\centering\arraybackslash}m{\mylength}}
&Map \#1 &Map \#2 &Map \#3\\
%
\rotatebox{90}{\Fullb}
\forloop{ct}{1}{\value{ct} < 4}{&\includegraphics[width=\mylengths]{images/4.RealData/Maps/Full10ms_\arabic{ct}.png} }\\
\rotatebox{90}{\Fulla}
\forloop{ct}{1}{\value{ct} < 4}{&\includegraphics[width=\mylengths]{images/4.RealData/Maps/Full2ms_\arabic{ct}.png} }\\
\rotatebox{90}{\deFull}
\forloop{ct}{1}{\value{ct} < 4}{&\includegraphics[width=\mylengths]{images/4.RealData/Maps/Denoised_\arabic{ct}.png} }\\

\rotatebox{90}{\SNN}
\forloop{ct}{1}{\value{ct} < 4}{&\includegraphics[width=\mylengths]{images/4.RealData/Maps/SNN_\arabic{ct}.png} }\\
\rotatebox{90}{\SSS}
\forloop{ct}{1}{\value{ct} < 4}{&\includegraphics[width=\mylengths]{images/4.RealData/Maps/SSS_\arabic{ct}.png} }\\
\end{tabular}
\egroup
\caption{Abundance maps estimated by SUNSAL.}
\label{Figure13b:Maps}
\end{figure}

\begin{figure*}
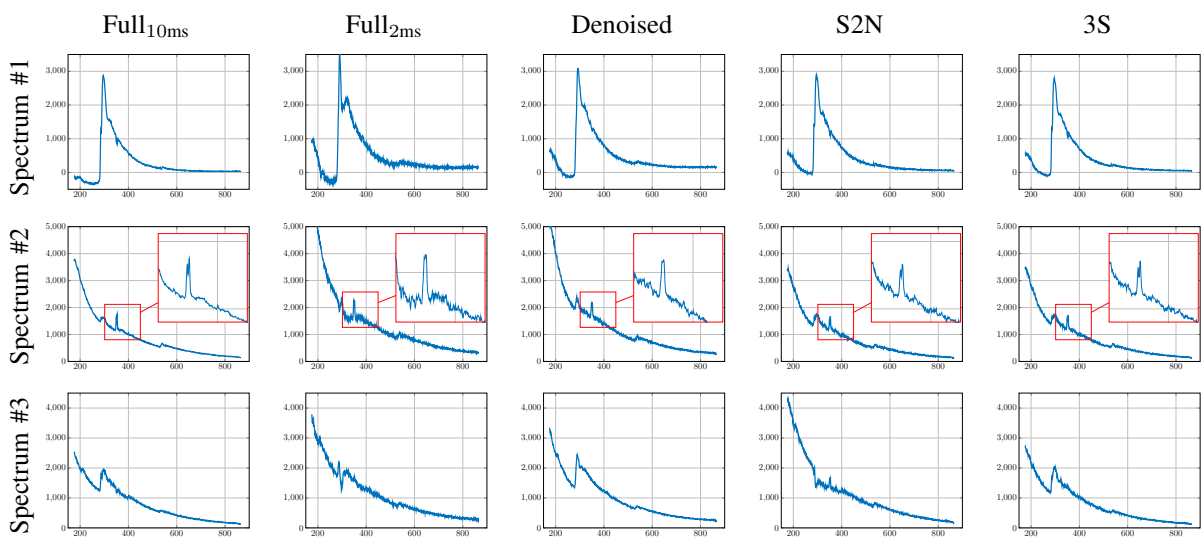

\centering
\setlength{\mylength}{4cm}
\setlength{\mylengths}{3.5cm}
\newcounter{cte}

\bgroup
\def\arraystretch{1.5}%
\begin{tabular}{>{\centering\arraybackslash}m{0.02cm}>{\centering\arraybackslash}m{\mylengthS}>{\centering\arraybackslash}m{\mylengthS}>{\centering\arraybackslash}m{\mylengthS}>{\centering\arraybackslash}m{\mylengthS}>{\centering\arraybackslash}m{\mylengthS}}
&\Fullb&\Fulla&\deFull&\SNN &\SSS \\
\rotatebox{90}{Spectrum \#1}
&\resizebox{\mySpectrafigurewidth}{!}{\input{images/4.RealData/Spectra/spectra_1_3.tex}}%
&\resizebox{\mySpectrafigurewidth}{!}{\input{images/4.RealData/Spectra/spectra_1_1.tex}}%
&\resizebox{\mySpectrafigurewidth}{!}{\input{images/4.RealData/Spectra/spectra_1_2.tex}}%
&\resizebox{\mySpectrafigurewidth}{!}{\input{images/4.RealData/Spectra/spectra_1_4.tex}}%
&\resizebox{\mySpectrafigurewidth}{!}{\input{images/4.RealData/Spectra/spectra_1_5.tex}}%
\\
\rotatebox{90}{Spectrum \#2}
&\resizebox{\mySpectrafigurewidth}{!}{\input{images/4.RealData/Spectra/spectra_2_3.tex}}%
&\resizebox{\mySpectrafigurewidth}{!}{\input{images/4.RealData/Spectra/spectra_2_1.tex}}%
&\resizebox{\mySpectrafigurewidth}{!}{\input{images/4.RealData/Spectra/spectra_2_2.tex}}%
&\resizebox{\mySpectrafigurewidth}{!}{\input{images/4.RealData/Spectra/spectra_2_4.tex}}%
&\resizebox{\mySpectrafigurewidth}{!}{\input{images/4.RealData/Spectra/spectra_2_5.tex}}%
\\
\rotatebox{90}{Spectrum \#3}
&\resizebox{\mySpectrafigurewidth}{!}{\input{images/4.RealData/Spectra/spectra_3_3.tex}}%
&\resizebox{\mySpectrafigurewidth}{!}{\input{images/4.RealData/Spectra/spectra_3_1.tex}}%
&\resizebox{\mySpectrafigurewidth}{!}{\input{images/4.RealData/Spectra/spectra_3_2.tex}}%
&\resizebox{\mySpectrafigurewidth}{!}{\input{images/4.RealData/Spectra/spectra_3_4.tex}}%
&\resizebox{\mySpectrafigurewidth}{!}{\input{images/4.RealData/Spectra/spectra_3_5.tex}}%
\\
\end{tabular}
\egroup
\caption{Three (out of five) endmember spectra estimated by SISAL. The axes are amplitude vs. energy loss (in eV).}
\label{Figure13a:Spectra}
\end{figure*}

As in the previous section, to evaluate the performances of the proposed methods, these images are unmixed using SISAL to extract the  signatures of the materials and SUNSAL to estimate the corresponding abundance maps. The maps are depicted in Fig.~\ref{Figure13b:Maps} while the endmember signatures are given in Fig.~\ref{Figure13a:Spectra}. A number of $N_c=5$ components has been chosen but only the three most significant are displayed. Visually, acquiring and then denoising the image following protocol $\mathcal{P}_1$ seems to give the best results since the spectrum-image and the abundance maps show better spatial resolution, revealing for instance details in the calcification area. However the spectral features are better recovered with the partial sampling acquisition with $\Delta t=10$ms (i.e., under Protocol $\mathcal{P}_2$). Indeed, the splitting on the Ca $L_{2,3}$ edge is clearly visible on the spectra $\sharp 2$ extracted from the reconstructed spectrum-images after protocol $\mathcal{P}_2$ (see zoomed areas in Fig.~\ref{Figure13a:Spectra}). However, this $L_{2,3}$ splitting is not resolved for the spectrum $\sharp 2$ extracted from the denoised image after protocol $\mathcal{P}_1$. 
Thought, it should be noted that in this case, the low number of pixels ($51\times51$) compared to the simulated data set in Section \ref{SectionSimulation} ($100\times100$) makes the extraction of components by SISAL less efficient. Hence, the choice of the acquisition strategy must result of a compromise between the required spatial and spectral resolutions.

\section{Conclusion}\label{SectionConclusion}

In this paper, we introduced new acquisition and reconstruction techniques to better preserve sensitive materials in transmission electron microscopy. 
The proposed methods are based on a partial acquisition of an EELS spectrum-image followed by reconstruction using a priori information. Two algorithms were proposed to conduct the reconstruction task and experiments compared this approach with a standard acquisition scheme. The results showed that the \SSS\ algorithm performs better than \SNN\, both in terms of quality of the reconstruction and computation time. 

When comparing with a standard (full) acquisition followed by denoising, the partial sampling scheme showed better spectra estimation, while some spatial details seemed deteriorated. 
Note that there are other benefits of partial sampling, including a better distribution of the energy within the sample, and the ability of reconstructing dynamic (temporal) sequences.
This opens new perspectives towards fast or dynamic STEM-EELS imaging.

\appendices

\section{Tuning the \SNN\ regularization parameters}\label{SubSec:empirical-search}
\label{app:adjusting_SNN_parameters}

This paragraph discusses the choice of the regularization parameters $\lSNN$ and $\mSNN$ adjusting the spatial and spectral regularizations in the \SNN\ objective function \eqref{eq:SNN}. To properly adjust the pair of parameters\footnote{To lighten the notations, the subscripts \SNN\ are omitted in the sequel of this section.} $(\lambda,\mu)$, and by denoting $\Y_{\indb,\I(n)}$ the $\indb$th component of the spectra $\Y_{\I(n)}$ measured at the spatial position indexed by $\I(n)$, the proposed strategy relies on the assumption
\begin{equation}
 \mathrm{E}\left[\left(\Y_{\indb,\I(n)}-\tX_{\indb,\I(n)}\right)^2\right] = \sigma^2
\end{equation}
which relates the noise variance and the expected reconstruction error in each energy band and for each pixel. Based on this assumption, choosing the optimal solution $\hat{\tX}^{\textrm{opt}} \triangleq \hat{\tX}\left(\lambda^{\textrm{opt}},\mu^{\textrm{opt}}\right)$ among the set of solutions $\left\{\hat{\tX}{\left(\lambda,\mu\right)}\right\}_{\lambda,\mu}$ would consist in solving the problem
\begin{equation}
 \left(\lambda^{\textrm{opt}},\mu^{\textrm{opt}}\right) \in \operatornamewithlimits{argmin}_{(\lambda,\mu)\in\mathbb{R}_+^2} \mathcal{J}\left(\lambda,\mu\right)
 \end{equation}
with  
\begin{equation}
    \mathcal{J}\left(\lambda,\mu\right) \triangleq \left(\frac{1}{\Ns\Nb}\left\|\Y_{\I}-\hat{\tX}_{\I}{(\lambda,\mu)}\right\|_{\mathrm{F}}^2-\hat{\sigma}^2\right)^2
\end{equation}
where $\hat{\sigma}^2$ is an estimate of the noise variance (see Section \ref{SubSec:3S}). To overcome the inextricability of this problem, the regularization parameters $\left(\lambda,\mu\right)$ are set as
\begin{equation*}
\left\{
  \begin{array}{cc}
    \lambda^* &= c^\circ \lambda^{\circ} \\
    \mu^*     &= c^\circ \mu^{\circ}
  \end{array}
\right.
\end{equation*}
where $\lambda^{\circ}$, $\mu^{\circ}$ and $c^*$ are successively estimated by solving
\begin{align}
 \lambda^{\circ} &\in \operatornamewithlimits{argmin}_{\lambda\in\mathcal{G}_{\lambda}} \mathcal{J}(\lambda,0) \label{eq:adjust_lambdaSNN}\\
  \mu^{\circ} &\in \operatornamewithlimits{argmin}_{\mu\in\mathcal{G}_{\mu}} \mathcal{J}(0,\mu) \label{eq:adjust_muSNN}\\
  c^\circ &\in \operatornamewithlimits{argmin}_{c\in\mathcal{G}_c} \mathcal{J}(c\lambda^{\circ},c\mu^{\circ})\label{eq:adjust_cSNN}
 \end{align} 
on the respective search grids $\mathcal{G}_{\lambda}$, $\mathcal{G}_{\mu}$ and $\mathcal{G}_{c}$ dynamically adapted by dichotomy processes. The first two steps \eqref{eq:adjust_lambdaSNN} and \eqref{eq:adjust_muSNN} adjust independently the weights of the spatial and spectral regularizations in the \SNN\ cost function \eqref{eq:SNN} respectively. The third step \eqref{eq:adjust_cSNN} aims at rescaling the parameters  $\lambda^{\circ}$ and $\mu^{\circ}$ to reduce the impact of considering the spatial and spectral regularizations jointly while preserving their respective proportions in \eqref{eq:SNN}.

\section{Estimating the weights and the subspace dimension in \SSS}~\label{Appendix:Weights}
\begin{figure}[ht]
\centering
\includegraphics[width=0.8\columnwidth]{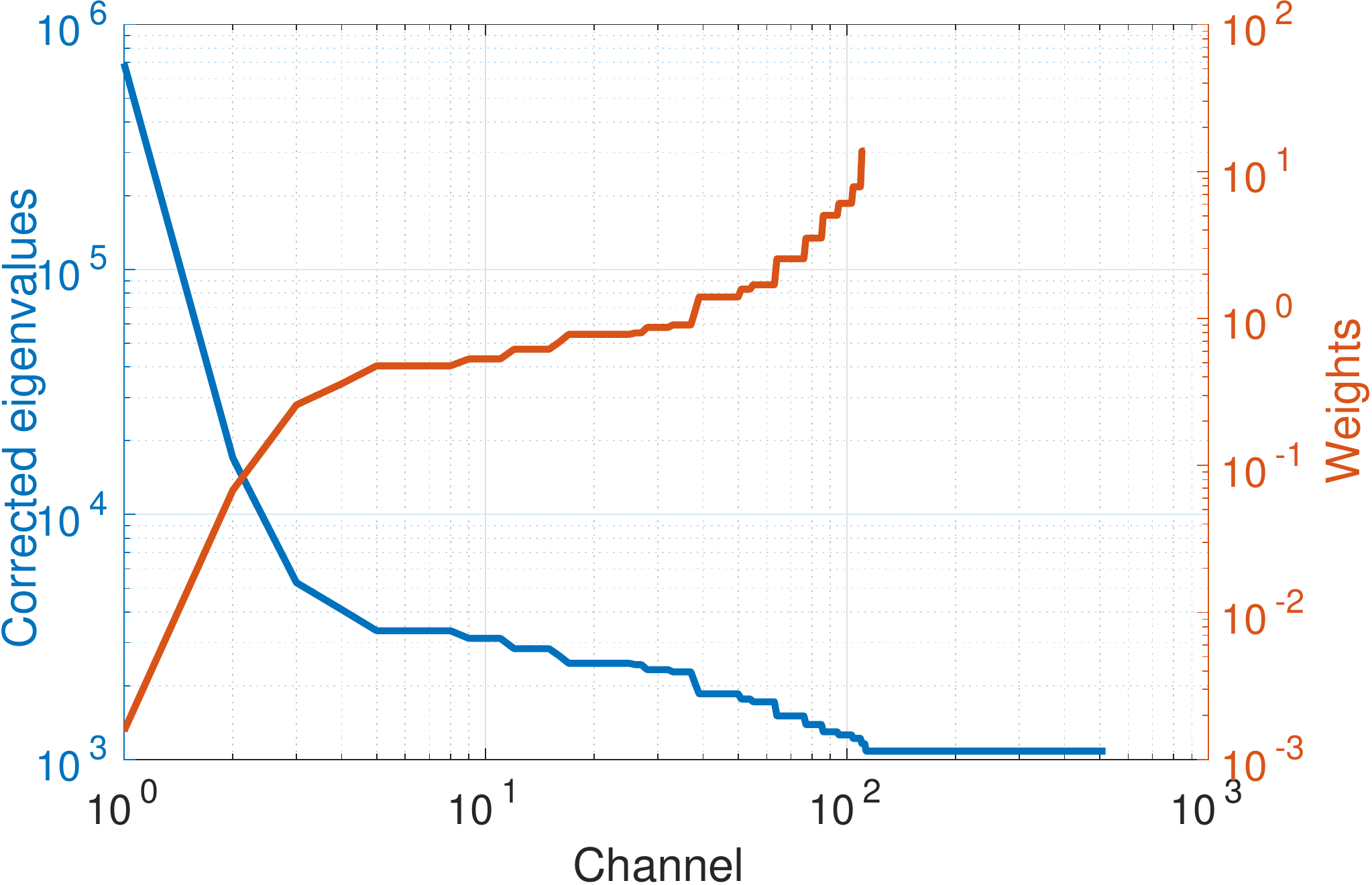}
\caption{A representation of the corrected eigenvalues (blue line) and the associated weights (in red). A real data spectrum image has been used here. Weights of index superior to 112 are infinity due to the equality between $\hat{\sigma}$ and $\hat{d}^2_{\Nb}$.}
\label{Fig4:S_constraint}
\end{figure}

As mentioned earlier, the vectors $\mathbf{h}_1,\ldots,\mathbf{h}_\rhoM$ spanning the subspace of interest are assumed to be sorted with respect to their corresponding eigenvalues $\d_1\geq\d_2\geq \ldots\geq \d_{\Nb}$ of decreasing magnitude. 
In particular, the latest directions are likely associated with noise. The representation vectors $\S_{\indb,:}$ ($\indb=1,\ldots,\Nb$) are thus also expected to be less relevant when $\indb$ increases. The choice of the weights $\w_{\indb}$ ($\indb=1,\ldots,\Nb$) is driven by this finding, by interpreting them within a Bayesian framework.
More precisely, let assume that the measurement matrix $\Y$ in case of full spatial sampling can be related to the unknown spectrum-image $\tX$ through the standard denoising model
\begin{equation}
    \Y = \tX + \mathbf{E}
\end{equation}
where $\mathbf{E}$ is a $\Nb \times \Ns$ noise matrix. This measurement equation can be reformulated in the signal subspace as
\begin{equation}
\H^T\Y = \S + \noise
\label{Equation:additiveNoise}
\end{equation}
with $\noise \triangleq \H^T\mathbf{E}$. In \eqref{Equation:additiveNoise}, $\noise=\left[\mathbf{N}_{1,:}^T,\ldots,\mathbf{N}_{\Nb,:}^T\right]^T$ denotes a perturbation matrix whose rows $\mathbf{N}_{\indb,:}$ ($\indb=1,\ldots,\Nb$) are assumed to be independent and identically distributed (i.i.d.) according to the normal distribution
\begin{equation}
    \mathbf{N}_{\indb,:} \sim \mathcal{N}\left(\boldsymbol{0}_{\Ns},\tsig^2 \boldsymbol{I}_{\Ns}\right).
\end{equation}
The rows $\S_{\indb,:}$ ($\indb=1,\ldots,\Nb$) of $\S$ are assumed to be i.i.d. and are assigned the following conjugate Gaussian prior
 \begin{equation}
{\S}_{\indb,:} \sim \mathcal{N}\left(\boldsymbol{0}_{\Ns},\eta^2_{\indb} \boldsymbol{I}_{\Ns}\right).
\end{equation}
Computing the maximum a posteriori (MAP) estimator of $\S$ consists in solving
\begin{equation}
\operatornamewithlimits{min}_{\S} \|\H^T\Y-\S\|_{\mathrm{F}}^2 + \sum_{\indb = 1}^\Nb \frac{\tsig^2}{\eta^2_{\indb}}\|{\S}_{\indb,:}\|_2^2.
\label{eq:3S_OP_2_Bayesian}
\end{equation}
By comparing the \SSS\ problem \eqref{eq:3S_OP_2} and the MAP formulation \eqref{eq:3S_OP_2_Bayesian}, a natural choice for the weights $w_{\indb}$ is
\begin{equation}
    w_{\indb} = \frac{\tsig^2}{\eta^2_{\indb}}.
\end{equation}
However, in practice the variance $\eta^2_{\indb}$ of the components ${\S}_{\indb,:}$ ($\indb=1,\ldots,\Nb$) are unknown. To adjust these hyperparameters, one solution consists in resorting to an empirical Bayesian approach by conducting a covariance analysis of the linear model \eqref{Equation:additiveNoise}, which straightforwardly leads to 
\begin{equation}
\d_{\indb} = \eta^2_{\indb} + \sigma^2
\end{equation}
where $\d_{\indb}$ ($\indb=1,\ldots,\Nb$) can be approximated by the sample eigenvalues $\hat{d}^2_{\indb}$ estimated by PCA after the correction detailed in Appendix~\ref{Appendix:Eigenvalues}. Finally, after defining the estimate $\hat{\sigma}^2$ of $\sigma^2$ as the corrected eigenvalue $\hat{d}^2_{\indb}$ of lowest magnitude (whose multiplicity order could be more than one), the weights are chosen as 
\begin{equation}
\label{eq:weight_definition}
\w_{\indb} = \frac{\hat{\tsig}^2}{\hat{d}^2_{\indb}-\hat{\tsig}^2}
\end{equation}
An illustration of the dependence between $\hat{d}^2_{\indb}$ and $\w_{\indb}$ is depicted in Fig. \ref{Fig4:S_constraint}. It is worth noting that, in the applicative context considered in this paper, the number of channels \Nb\ is usually of the same order of magnitude as \Ns. Thus the PCA conducted to estimate the eigenvalues $d_1,\ldots,d_\Nb$ may suffer from sample starvation and provide unreliable estimates. To improve this estimation, a correction of the PCA eigenvalue estimates is detailed in Appendix~\ref{Appendix:Eigenvalues}.

Moreover, the rule \eqref{eq:weight_definition} also suggests to define an estimate $\rhoM$ of the signal subspace dimension $\rhoM_{\mathrm{true}}$ as the maximum index $\indb$ such that $\hat{d}^2_{\indb+1}>\hat{\tsig}^2$. For $\indb=\rhoM+1,\ldots,\Nb$, the weights are set as $\w_{\indb}=\infty$ since the corresponding representation vectors are expected to be composed of noise only. Hence, these components ${\S}_{\indb,:}$ ($\indb=\rhoM+1,\ldots,\Nb$) are enforced to be null and the \SSS\ optimization problem \eqref{eq:3S_OP_2} can be reformulated to be minimized with respect to a $\rhoM\times\Np$ matrix. 

\section{Correcting the eigenvalues estimated by PCA}\label{Appendix:Eigenvalues}

The rule detailed in Appendix \ref{Appendix:Weights} to adjust the \SSS\ weights requires an estimation of the variances $d_1^2,\ldots,d_{\Nb}^2$ of the signal components in the basis spanned by $\H$. When $\H$ is identified by PCA, this estimation is classically performed by conducting an eigen-decomposition of the sample covariance matrix of the observed pixel spectra, i.e.,
\begin{equation}
    \hat{\Cov}=\frac{1}{\Ns}\Y_\I\Y_\I^T=\H\mathbf{G}\H^T
\end{equation}
where  $\mathbf{G}=\mathrm{diag}(\tilde{d}^2_1,\cdots,\tilde{d}^2_\Nb)$. Note here that the sample eigenvalues are sorted in a decreasing order and are positive, i.e., $\tilde{d}^2_1\geq\tilde{d}^2_2\geq\dots\geq\tilde{d}^2_\Nb\geq0$. The main drawback of this simple estimator is the fact that it is designed to provide good estimate when the sample size \Ns\ is sufficiently high compared to the observation dimension \Nb. However, this is not the case in the considered applicative context since \Ns\ is of the same order of magnitude as \Nb. Various strategies have been proposed in the literature to improve eigenvalue estimation. To correct the sample eigenvalues, one alternative consists in resorting to the Stein estimator defined as \cite{mestre2008improved}
\begin{equation}
    \hat{d}^2_\indb=\frac{\tilde{d}^2_\indb}{1+ \frac{1}{\Ns}\sum_{\substack{j=1\\j\neq \indb }}^\Nb \frac{\tilde{d}^2_\indb+\tilde{d}^2_j}{\tilde{d}^2_\indb-\tilde{d}^2_j}}.
\end{equation}
However, this estimator does not ensure the non-increasing property and some corrected eigenvalues $\hat{d}^2_\indb$ ($\indb=1,\ldots,\Nb$) can be negative. To alleviate this issue, an isotonic regression has been proposed in \cite{LinPerl1985} as a post-processing step. This procedure usually returns a set of corrected eigenvalues with associated multiplicity orders. Moreover, an estimate $\hat{\sigma}^2$ of the noise variance $\sigma^2$ required in the weight definition \eqref{eq:weight_definition} can be chosen as the corrected eigenvalues of lowest magnitude, whose multiplicity order is expected to be $\Nb-\rhoM$.

\bibliographystyle{IEEEtran}
\bibliography{\biblio}

\end{document}